\documentclass[twocolumn,floatfix,preprintnumbers,nofootinbib,superscriptaddress]{revtex4}

\usepackage{ulem}
\usepackage{bm}
\usepackage{times}
\usepackage{amssymb,amsbsy,amsmath,amsfonts}
\usepackage{graphicx}
\usepackage{float}
\usepackage{color}
\usepackage{morefloats}
\usepackage{rotating}
\usepackage{srcltx}
\usepackage{slashed}
\usepackage{multirow}
\usepackage{verbatim}
\usepackage{hyperref}
\usepackage{tabularx}
\usepackage{braket}
\usepackage{amsfonts}
\usepackage{color}
\usepackage{mathrsfs}
\usepackage{gensymb}
\usepackage{booktabs}  
\usepackage{threeparttable}  
\usepackage[outdir=./]{epstopdf}

\newcommand\be{\begin{eqnarray}}
\newcommand\ee{\end{eqnarray}}

\begin{document}

\title{The electromagnetic form factors of $\Sigma$ and $\Sigma^0 \to \Lambda$ transition in the timelike region}

\author{Cheng Chen}
\affiliation{State Key Laboratory of Heavy Ion Science and Technology, Institute of Modern Physics, Chinese Academy of Sciences, Lanzhou 730000, China}
\affiliation{School of Nuclear Sciences and Technology, University of Chinese Academy of Sciences, Beijing 101408, China}

\author{Bing Yan}
\affiliation{College of Physics, Sichuan University, Chengdu, Sichuan 610065, China}

\author{Bing-Wei Long}
\affiliation{College of Physics, Sichuan University, Chengdu, Sichuan 610065, China}
\affiliation{Southern Center for Nuclear-Science Theory (SCNT), Institute of Modern Physics, Chinese Academy of Sciences, Huizhou 516000, China}

\author{Ju-Jun Xie}~\email{xiejujun@impcas.ac.cn}
\affiliation{State Key Laboratory of Heavy Ion Science and Technology, Institute of Modern Physics, Chinese Academy of Sciences, Lanzhou 730000, China} \affiliation{School of Nuclear Sciences and Technology, University of Chinese Academy of Sciences, Beijing 101408, China} \affiliation{Southern Center for Nuclear-Science Theory (SCNT), Institute of Modern Physics, Chinese Academy of Sciences, Huizhou 516000, China}

\begin{abstract}

We investigate the $e^+e^-\to \Sigma\bar{\Sigma}$ and $e^+e^-\to \Lambda\bar{\Sigma}^0$ reactions within the extended vector meson dominance model. In addition to the ground state mesons $\rho$ and $\omega$, we consider the contributions of the excited states $\rho(3D)$, $\omega(3D)$, $\phi(3D)$, and $\rho(6D)$. It is found that the current experimental data on the $\Sigma$ electromagnetic form factors in timelike region can be well reproduced. And the $\phi(3D)$ resonance is essential to the near threshold enhancement of the cross section for the $e^+e^-\to \Sigma\bar{\Sigma}$ reaction. Furthermore, in the $e^+e^-\to \Lambda\bar{\Sigma}^0$ reaction, the $\rho(3D)$ is important to get a good fit for the experimental results.

\end{abstract}

\maketitle
\section{Introduction}

The electromagnetic form factors (EMFFs) of baryons encode crucial information about their electromagnetic structure~\cite{Denig:2012by,Pacetti:2014jai,Punjabi:2015bba}, which depends on the squared of four-momentum transfer, $q^2$, of the interacted virtual photon. Hence, investigating the EMFFs of baryons is essential for gaining deeper insight into their fundamental structure. The region where $q^2 < 0$ is referred to as the spacelike region, while $q^2 > 0$ is known as the timelike region. In the spacelike region, EMFFs are real and relate to the charge and magnetic distribution of baryons. In contrast, EMFFs in the timelike region are generally complex. Interestingly, an analytical study revealed that EMFFs in the timelike region tend to become real at large four-momentum transfer squared~\cite{Mangoni:2021qmd}. Furthermore, Ref.~\cite{Kuraev:2011vq} proposed that, in the timelike region, these EMFFs are connected to the time evolution of baryons' electromagnetic distributions. On the experimental side, measuring the EMFFs of unstable baryons, such as hyperon $\Sigma$, $\Lambda$, and $\Xi$, through electron-baryon scattering in the spacelike region is challenging due to their short lifetimes. Consequently, it is more practical to study baryon EMFFs in the timelike region using electron-positron annihilation reactions. Significant experimental progress has been made in measuring hyperon EMFFs in the timelike region~\cite{BaBar:2007fsu,BESIII:2017hyw,BESIII:2019cuv,BESIII:2019nep,BESIII:2021aer,BESIII:2021rkn,BESIII:2021ccp,Belle:2022dvb,BESIII:2023rse,BESIII:2023ldb,BESIII:2023pfv,BESIII:2024umc}. Recent advancements in experimental techniques and increased statistics have considerably reduced measurement uncertainties. Additionally, spin polarization distributions of final-state baryons have been obtained in reactions such as $e^+e^- \to \Lambda \bar{\Lambda}$~\cite{BESIII:2018cnd,BESIII:2021cvv,BESIII:2022qax,BESIII:2023euh}, $e^+e^- \to \Sigma \bar{\Sigma}$~\cite{BESIII:2020fqg,BESIII:2023ynq,BESIII:2024nif,BESIII:2024dmr} and $e^+e^- \to \Xi \bar{\Xi}$~\cite{BESIII:2022lsz,BESIII:2023drj,BESIII:2023lkg,Liu:2023xhg}. These rich experimental data have sparked extensive theoretical interest in studying EMFFs of the hyperons and related topics~\cite{Cao:2018kos,Yang:2019mzq,Haidenbauer:2020wyp,Li:2020lsb,Li:2021lvs,Dai:2021yqr,Yan:2023yff,Lin:2022baj,Dai:2023vsw,Lin:2023qnv,Cao:2024tvz,Zhang:2024rbl,Zheng:2025tnz,Dai:2024lau}.

In this study, we focus on the $e^+e^- \to \Sigma \bar{\Sigma}$ and $e^+e^- \to \Lambda \bar{\Sigma}^0$ processes. First, for the $e^+e^- \to \Sigma \bar{\Sigma}$ process, experimental measurements from the BESIII Collaboration reveal that the total cross-section ratios for the reactions $e^+e^- \to \Sigma^+ \bar{\Sigma}^-$, $\Sigma^0 \bar{\Sigma}^0$, and $\Sigma^- \bar{\Sigma}^+$ are $9.7 \pm 1.3:3.3 \pm 0.7:1$ over a center-of-mass energy range of 2.3864 GeV to 3.02 GeV~\cite{BESIII:2020uqk,Mangoni:2021lmr,Irshad:2022zga}. These ratios deviate significantly from predictions made by various theoretical models~\cite{Ramalho:2019koj,Anselmino:1992vg,Kubis:2000aa}. The discrepancy has drawn considerable attention, prompting theoretical investigations from different perspectives. The vector meson dominance (VMD) model, a well-established framework for studying nucleon electromagnetic form factors in both space and timelike regions~\cite{Iachello:1972nu,Iachello:2004aq,Bijker:2004yu,Yan:2023nlb}, has also been extended to the hyperon sector~\cite{Li:2020lsb,Li:2021lvs,Dai:2021yqr,Yan:2023yff}, In this model, the virtual photon couples to baryons via intermediate vector mesons. In our earlier work, we applied the VMD model to the $e^+e^- \to \Sigma \bar{\Sigma}$ process, considering only the ground states of vector mesons. This approach successfully reproduced the observed cross-section ratios~\cite{Yan:2023yff}. However, recent results from BESIII for the $e^+e^- \to \Sigma^+ \bar{\Sigma}^-$ reaction indicate higher central values than earlier 2021 measurements. Additionally, the findings suggest the potential existence of a new resonance at $\sqrt{s} = 2.5 \, \mathrm{GeV}$~\cite{BESIII:2023ldb}. These findings suggest that including only the ground of vector mesons is insufficient to fully describe current experimental data. To address this, contributions excited states of the vector mesons must be incorporated into the model.

For the $e^+e^-\to \Lambda\bar{\Sigma}^0$ reaction, its total cross sections were measured for the first time by BaBar Collaboration~\cite{BaBar:2007fsu}, and it was recently re-measured by the BESIII Collaboration with improved precision~\cite{BESIII:2023pfv}. The two groups of experimental data are similar expect the one closed to the reaction threshold. The BESIII Collaboration obtained the total cross section $72.9\pm 12.6\pm 5.4$ pb with a statistical significance greater than 5 standard deviations with the center-of-mass energy just about 1 MeV above the threshold. And the fit results given at 2.386 GeV seem to exceed the upper limit of 75 pb set by DM2 Collaboration \cite{DM2:1990tut}. In Ref.~\cite{Haidenbauer:2020wyp}, the author used various $\Lambda\bar{\Sigma}^0$ potential form \cite{Haidenbauer:1993ws} as final state interaction reproduced the reaction of the near-threshold $e^+e^-\to \Lambda\bar{\Sigma}^0$ cross section measured by BarBar Collaboration. Their result showed a sudden increase in the cross-section near the threshold, followed by a plateau extending over 100 MeV. However, this behavior is inconsistent with the latest experimental data from the BESIII Collaboration~\cite{BESIII:2023pfv}, suggesting that additional mechanisms must be taken into account. 

For excited states of vector mesons, numerous experimental results are documented in the Particle Data Group (PDG)~\cite{ParticleDataGroup:2024cfk}. However, experimental discoveries for light vector mesons with masses above 2 GeV remain limited. As a result, most studies on highly excited vector mesons rely on theoretical approaches~\cite{Pang:2019ovr,Wang:2021gle,Wang:2021abg,Feng:2021igh,Zhou:2022wwk,Zhou:2022ark,Wang:2022xxi,Wang:2022juf,Bai:2023dhc,Lodha:2024qby,Lodha:2024bwn}. In Ref.~\cite{Wang:2021gle}, the excited states of $\rho$ and $\omega$ mesons were explored within the 2.0–2.4 GeV mass range. The $\rho(3D)$ and $\omega(3D)$ states were identified with masses of 2.283 GeV and widths of 158 MeV and 94 MeV, respectively, both lying close to the $\Sigma\bar{\Sigma}$ threshold. In Ref.~\cite{Wang:2021abg}, it investigated the mass and decay properties of $\rho$, $\omega$, and $\phi$ meson excited states in the 2.4–3.0 GeV range using the modified Godfrey-Isgur model, identifying six excited states for each meson within this interval. And the studies in Ref.~\cite{Feng:2022hwq} expanded on the decay properties of highly excited rho states, considering additional decay channels beyond those in~\cite{Wang:2021abg}. In addition, the decay of $\phi$ excited states into $\Lambda\bar{\Lambda}$, $\Sigma\bar{\Sigma}$, and $\Xi\bar{\Xi}$ was also examined by using the hadronic-loop mechanism~\cite{Bai:2023dhc}. It was found that the $\phi(3D)$ state at 2.5 GeV decays into $\Sigma\bar{\Sigma}$ with a relatively significant branching ratio, suggesting its potential importance in the $e^+e^- \to \Sigma\bar{\Sigma}$ reaction.

Based on these theoretical predictions and the vector mesons listed in the PDG~\cite{ParticleDataGroup:2024cfk}, we include the $\rho(3D)$, $\omega(3D)$, $\phi(3D)$, and $\rho(6D)$ states for studying the $e^+e^- \to \Sigma\bar{\Sigma}$ and $e^+e^- \to \Lambda \bar{\Sigma}^0$ reactions within the framework of the VMD model. The $\rho(6D)$ state was introduced to account for experimental data at a center-of-mass energy of 2.9 GeV, with its mass and width set to 2.850 GeV and 150 MeV, respectively. Their masses and widths are summarized in Table~\ref{tab:massandwidth}.

\begin{table}[htbp]
    \centering
 \caption{The mass and width of the states used for this work. We take an averaged mass and width for $\omega$ and $\phi$ mesons.}
    \begin{tabular}{c | c | c |c}\hline\hline
        State & Mass $M_{R}$ (MeV) & Width $\Gamma_R$ (MeV) & Reference \\\hline
        $\rho$ & $775$ & $149.1$ & \cite{ParticleDataGroup:2024cfk}  \\
        $\omega\phi$ & $900.5$ & $6.46$ & ~\cite{ParticleDataGroup:2024cfk} \\
        $\rho(3D)$ & $2300$ & $160$ & ~\cite{Wang:2021gle}\\
        $\omega(3D)$ & $2300$ & $100$ & ~\cite{Wang:2021gle} \\
        $\phi(3D)$ & $2500$ & $170$ & ~\cite{Bai:2023dhc,Wang:2021abg}\\
        $\rho(6D)$ & $2850$ & $150$ & ~\cite{Wang:2021abg,Feng:2022hwq}\\
        \hline\hline
    \end{tabular}
       \label{tab:massandwidth}
\end{table}

This article is organized as follows: in the following section, we outline the theoretical framework for calculating the EMFFs of the $\Sigma$ and $\Lambda$-$\bar{\Sigma}^0$ transition within the VMD approach. Numerical results and discussions are presented in Sec.~\ref{sec:numerical results}.  Finally, a brief summary and conclusions are provided in the last section.

\section{Formalism}  \label{sec:formalism}

For the $e^+e^-\to \Sigma\bar{\Sigma}$ reaction, the interaction vertex of $\gamma^* \Sigma \bar{\Sigma}$ can be parameterized to two independent form factors $F_1(q^2)$ and $F_2(q^2)$, were called the Dirac and Pauli form factors, respectively, according to the electromagnetic current conservation and $CP$ invariance. Therefore, the matrix of $\Sigma\bar{\Sigma}$ created by electromagnetic current from the vacuum can be expressed as
\begin{eqnarray}
&& \!\!\!\!\!\! \braket{\Sigma\bar{\Sigma}|J_\mu^{em}|0} = \nonumber \\
&& \!\!\!\!\!\! \bar{u}_\Sigma(p_\Sigma,s_\Sigma) \left[ \gamma_\mu F_1(q^2) + i\frac{F_2(q^2)}{2M_\Sigma}\sigma_{\mu\nu}q^\nu \right] v_{\bar{\Sigma}} (p_{\bar{\Sigma}},s_{\bar{\Sigma}}), \label{eq:emc}
\end{eqnarray}
where $q$ is four-momentum of the virtual photon $\gamma^*$, and $M_{\Sigma}$ is the mass of $\Sigma$ hyperon. Then the electric $G_E$ and magnetic form factor $G_M$ can be obtained by combining the Pauli and Dirac form factors,
\begin{eqnarray}
    G_E(q^2) &=& F_1(q^2) + \tau F_2(q^2), \label{eq:ge} \\
    G_M(q^2) &=& F_1(q^2) +  F_2(q^2), \label{eq:gm}
\end{eqnarray}
with $\tau = q^2/(4M_{\Sigma}^2)$. Within the $G_M$ and $G_E$, the total cross section of the $e^+e^- \to \Sigma \bar{\Sigma}$ reaction in the center of mass frame is obtained as
\begin{align}
    \sigma(e^+e^-\to\Sigma\bar{\Sigma}) = \frac{4\pi\alpha_{em}^2\beta C}{3s}\left( |G_M|^2 + \frac{1}{2\tau}|G_E|^2 \right), \label{eq:cs1}
\end{align}
where $s=q^2$ is the invariant mass squared of the $e^+ e^-$ system, $\alpha_{em}=e^2/4\pi=1/137.036$ is the electromagnetic fine structure constant, and $\beta = \sqrt{1-4M^2_{\Sigma}/s}$ is the velocity of $\Sigma$ hyperon.
The factor $C$ is the $S$-wave Sommerfeld-Gamow factor originating from the Coulomb interaction of the final $\Sigma\bar{\Sigma}$~\cite{Sakharov:1948plh},
\begin{eqnarray}
    C=\begin{cases}
    \frac{y}{1-e^{-y}}, &\text{for}\  \Sigma^+, \Sigma^-,\\  
    & \\
    1,           &\text{for}\  \Sigma^0,
    \end{cases}
\end{eqnarray}
and here $y = 2\pi\alpha_{em}M_{\Sigma}/(\beta\sqrt{s})$. Thus, for charged $\Sigma^+$ and $\Sigma^-$, the total cross section of their responding is non-zero at the reaction threshold. Moreover, the Coulomb interaction just affects the reaction threshold about a few MeV, and it decreases rapidly and exponentially to a constant 1.

According to the total cross section as in Eq.~\eqref{eq:cs1}, we can extract the so-called effective form factor $|G_{\rm eff}|$ that is
\begin{eqnarray}
    | G_{\rm eff} (q^2) | = \sqrt{\frac{2\tau|G_M(q^2)|^2 + | G_E(q^2)|^2}{1+2\tau}}.
\end{eqnarray}
Within the one-photon exchange approximation, the effective form factor stands for the effective coupling strength between photon and hyperons, varying with the center mass of energy, and also characterizes the degree of $\Sigma$ hyperons deviation from the point particle. Moreover, we can also obtain the differential cross section distribution parameter $\alpha$ in the reaction
\begin{align}
    \alpha = \frac{s|G_M|^2 - 4M_\Sigma^2|G_E|^2}{s|G_M|^2 + 4M_\Sigma^2|G_E|^2},
\end{align}
which only depends on the ratio of the absolute values of EMFFs and satisfies $-1\leq \alpha \leq 1$.

In the timelike region, $G_E$ and $G_M$ are complex, and their relative phase $\Delta \Phi$ is an observable quantity, which can be extracted from the spin polarization of the final 
hyperons~\cite{BESIII:2018cnd,BESIII:2021cvv,BESIII:2022qax,BESIII:2023euh,BESIII:2020fqg,BESIII:2023ynq,BESIII:2024nif,BESIII:2024dmr,BESIII:2022lsz,BESIII:2023lkg,Liu:2023xhg}
\begin{align}
    G_E/G_M = |G_E/G_M|e^{i\Delta \Phi},
\end{align}
and the spin polarization distribution along the direction $\hat{p}\times\hat{k}$ ($\hat{k}$ is the momentum direction of electron, and $\hat{p}$ is the momentum direction of $\Sigma$ hyperon in the center mass frame) is given by~\cite{Faldt:2017kgy}
\begin{align}
    P_y(\theta) = \frac{\sqrt{1-\alpha^2}\sin \theta \cos\theta }{1+\alpha\cos^2\theta}\sin\Delta \Phi, \label{eq:pol}
\end{align}
where the $\theta$ is the scattering angle between the electron and $\Sigma$ hyperon. At a fixed center-of-mass energy, the maximum value of the polarization $P_y$ occurs when $\cos^2 \theta = 1/(2+\alpha)$, with ${P_y}^{max} = \frac{1}{2}\sqrt{1-\alpha}\sin\Delta \Phi$. Further details regarding polarization observables in the timelike region are provided in the Appendix.

For the $e^+e^-\to \Lambda\bar{\Sigma}^0$ reaction, the matrix of the $\Lambda \bar{\Sigma}^0$ created by the electromagnetic current is given by~\cite{Granados:2017cib} 
\begin{eqnarray}
    \braket{\Lambda \bar{\Sigma}^0|J_\mu^{em}|0}& =& \bar{u}_\Lambda (p_\Lambda,s_\Lambda) \left[ \left( \gamma_\mu - \frac{M_\Lambda - M_{\bar{\Sigma}^0}}{q^2}q^\mu \right)  F_1(q^2) \right. \nonumber \\ 
  &&\left. + i\frac{F_2(q^2)}{M_{\bar{\Sigma}^0}+M_\Lambda}\sigma_{\mu\nu}q^\nu \right] v_{\bar{\Sigma}^0}(p_{\bar{\Sigma}^0},s_{\bar{\Sigma}^0}).
\end{eqnarray}
Similarly, the total cross section of the reaction is 
\begin{eqnarray}
    \sigma(e^+e^-\to\Lambda\bar{\Sigma}^0) = \frac{4\pi\alpha_{em}^2\beta}{3s}\left( |G_M|^2 + \frac{1}{2\tau}|G_E|^2 \right), \label{eq:cs2}
\end{eqnarray}
with $ \tau = q^2/(M_{\Sigma^0}+M_\Lambda)^2$, and $\beta$ is given by
\begin{eqnarray}
    \beta = \sqrt{1-\frac{2(M_{\bar{\Sigma}^0}^2 + M_\Lambda^2)}{q^2} + \left(\frac{M_{\bar{\Sigma}^0}^2 - M_\Lambda^2}{q^2} \right)^2}.
\end{eqnarray}

In our previous work~\cite{Yan:2023yff}, only ground vector mesons $\rho$, $\omega$, and $\phi$ were included in the reaction $e^+e^-\to\Sigma\bar{\Sigma}$. Here, we will also include the excited vector states around 2 GeV. Since $\rho(3D)$ and $\omega(3D)$ have the same mass, it is difficult to determine the contribution of each of $\rho(3D)$ and $\omega(3D)$ if both are included in the VMD model. Therefore to analyse the effect of $\rho(3D)$ and $\omega(3D)$ on the form factors, and also to reduce the parameters, we considered two scenarios in the model: $\rho$, $\omega$, $\phi$, $\omega(3D)$, $\phi(3D)$ and $\rho(6D)$ are considered in Scenario I; while $\omega(3D)$ is replaced by $\rho(3D)$ in Scenario II.

For Scenario I, the Dirac and Pauli form factors $F_1$ and $F_2$ for $\Sigma$ isospin triplet states in the timelike region can be parametrized as
\begin{eqnarray}
    F_{1}^{\Sigma^{\pm}}&=&g(q^2)\left( f_{1}^{\Sigma^{\pm}} \pm \frac{\beta _{\rho}}{\sqrt{2}}B_{\rho}+\frac{\beta _{\omega \phi}}{\sqrt{3}}B_{\omega \phi} \right. \nonumber \\  
      && \left. +  \frac{\beta _{\omega(3D)}}{\sqrt{3}}B_{\omega(3D)} +\frac{\beta _{\phi(3D)}}{\sqrt{3}}B_{\phi(3D)} \right. \nonumber \\ 
    && \left. \pm \frac{\beta _{\rho(6D)}}{\sqrt{2}}B_{\rho(6D)}\right),  \\ 
    F_{2}^{\Sigma^{\pm}}&=&g(q^2)\left( f_{2}^{\Sigma^{\pm}}B_{\rho}+\frac{\alpha _{\omega \phi}}{\sqrt{3}}B_{\omega \phi}+\frac{\alpha_{\omega(3D)}}{\sqrt{3}}B_{\omega(3D)}\right. \nonumber \\
    && \left. +\frac{\alpha _{\phi(3D)}}{\sqrt{3}}B_{\phi(3D)} \pm \frac{\alpha _{\rho(6D)}}{\sqrt{2}}B_{\rho(6D)}\right) ,
 \end{eqnarray}
 for $\Sigma^+$ and $\Sigma^-$, and
 \begin{eqnarray}
    F_{1}^{\Sigma ^0}&=&g(q^2)\left( f_{1}^{\Sigma ^0}+\frac{\beta _{\omega \phi}}{\sqrt{3}}B_{\omega\phi}\right. \nonumber \\ 
    && \left.+\frac{\beta _{\omega(3D)}}{\sqrt{3}}B_{\omega(3D)}+\frac{\beta _{\phi(3D)}}{\sqrt{3}}B_{\phi(3D)} \right) ,
    \\
    F_{2}^{\Sigma ^0}&=&g(q^2)\left( f_{2}^{\Sigma ^0}B_{\omega \phi} +\frac{\alpha _{\omega(3D)}}{\sqrt{3}}B_{\omega(3D)} \right. \nonumber \\ 
    && \left.+\frac{\alpha _{\phi(3D)}}{\sqrt{3}}B_{\phi(3D)} \right) ,
 \end{eqnarray}
 for $\Sigma^0$, with
\begin{eqnarray}
    B_R = \frac{M_R^2}{M_R^2-q^2-i M_R\Gamma_R}, \label{eq:bw}
\end{eqnarray}
where the label $R$ denotes as these states $\rho$, $\omega \phi$, $\omega(3D)$, $\rho(3D)$, $\phi(3D)$ and $\rho(6D)$. 

At $q^2=0$ and setting width $\Gamma_R = 0$, the electric $G_E$ and magnetic $G_M$ should be equal to the charge and magnetic momentum of $\Sigma$ hyperons, respectively, i.e. $G_E^{\Sigma^{+}}=1$ and $G_M^{\Sigma^{+}}=\mu_{\Sigma^{+}}$, $G_E^{\Sigma^{0}}=0$ and $G_M^{\Sigma^{0}}=\mu_{\Sigma^{0}}$, $G_E^{\Sigma^{-}}=-1$ and $G_M^{\Sigma^{-}}=\mu_{\Sigma^{-}}$, respectively, and then the coefficients $f_1^{\Sigma^{+}}$ and $f_2^{\Sigma^{+}}$, $f_1^{\Sigma^{0}}$ and $f_2^{\Sigma^{0}}$, $f_1^{\Sigma^{-}}$ and $f_2^{\Sigma^{-}}$ can be determined by these constrains,
\begin{eqnarray}
f_1^{\Sigma^{\pm}}   &=&   \pm 1 -\frac{\beta _{\omega \phi} + \beta _{\omega(3D)} + \beta_{\phi(3D)}}{\sqrt{3}}  \nonumber \\
&&  \mp  \frac{\beta _{\rho} + \beta _{\rho(6D)}}{\sqrt{2}}, \\
    f_2^{\Sigma^{\pm}} &=& \mu_{\Sigma^{\pm}} \mp 1 -\frac{\alpha _{\omega \phi} + \alpha _{\omega(3D)} + \alpha_{\phi(3D)}}{\sqrt{3}} \nonumber \\
    && \mp \frac{\beta _{\rho(6D)}}{\sqrt{2}},
\end{eqnarray}
for $\Sigma^+$ and $\Sigma^-$, and
\begin{eqnarray}
    f_1^{\Sigma^{0}} &=& -\frac{\beta _{\omega \phi}}{\sqrt{3}} - \frac{\beta _{\omega(3D)}}{\sqrt{3}}-\frac{\beta_{\phi(3D)}}{\sqrt{3}}, \\
    f_2^{\Sigma^{0}} &=& \mu_{\Sigma^{0}} - \frac{\alpha _{\omega(3D)}}{\sqrt{3}}-\frac{\alpha_{\phi(3D)}}{\sqrt{3}}, \\ 
\end{eqnarray}
for $\Sigma^0$. Here, we take $\mu_{\Sigma^{+}}=3.112\hat{\mu}_{\Sigma^{+}}$, $\mu_{\Sigma^{-}}=-1.479\hat{\mu}_{\Sigma^{-}}$, and $\mu_{\Sigma^{0}}=2.044\hat{\mu}_{\Sigma^{0}}$ in natural unit $\hat{\mu}=\frac{e}{2M_{\Sigma}}$ ~\cite{ParticleDataGroup:2024cfk}. A more detailed derivation can be found in Ref.~\cite{Yan:2023yff}. Furthermore, corresponding expressions for Scenario \uppercase\expandafter{\romannumeral2} can easily be obtained just applying the substitution to $F_1$ and $F_2$ with  $\rho(3D) \to \omega(3D)$.


The function $g(q^2)$ represents a phenomenological intrinsic form factor, typically expressed in the dipole form as: 
\begin{align}
    g(q^2) = \frac{1}{(1-\gamma q^2)^2},
\end{align}
where $\gamma$ is a free parameter. It is worthy to note that, in the space-like region, substituting $Q^2 = -q^2$ into the dipole form of $g(Q^2)$, along with the corresponding expressions for $F_1(Q^2)$ and $F_2(Q^2)$, it is straightforward to derive their behavior for large $Q^2$. Specifically, for large $Q^2$, $F_1 \propto \frac{1}{Q^4}$ and $F_2 \propto \frac{1}{Q^6}$. These results are consistent with the asymptotic behavior of $F_1$ and $F_2$ predicted by perturbative quantum chromodynamics~\cite{Lepage:1979za,Lepage:1980fj,Belitsky:2002kj}.

For the case of $e^+e^-\to \Lambda\bar{\Sigma}^0$ reaction, since the total isospin of $\Lambda\bar{\Sigma}^0$ is 1, there are only vector mesons with isospin $I=1$ contributed to the process, in VMD model. So, we just consider two states $\rho$ and $\rho(3D)$. Therefore, the Dirac and Pauli transition form factor $F_1^{\Lambda\bar{\Sigma}^0}$ and $F_2^{\Lambda\bar{\Sigma}^0}$ can constructed as following
\begin{eqnarray}
    F_1^{\Lambda\bar{\Sigma}^0} &=& g(q^2)\left(f_1^{\Lambda\bar{\Sigma}^0} + \beta_\rho^{\Lambda\bar{\Sigma}^0} B_\rho + \beta_{\rho(3D)}^{\Lambda\bar{\Sigma}^0}B_{\rho(3D)}    \right) , \\
    F_2^{\Lambda\bar{\Sigma}^0} &=& g(q^2)\left(f_2^{\Lambda\bar{\Sigma}^0}B_\rho + \alpha_{\rho(3D)}^{\Lambda\bar{\Sigma}^0}B_{\rho(3D)}    \right),
\end{eqnarray}
with
\begin{eqnarray}
    f_1^{\Lambda\bar{\Sigma}^0} &=& -\beta_\rho^{\Lambda\bar{\Sigma}^0} - \beta_{\rho(3D)}^{\Lambda\bar{\Sigma}^0} ,\\
    f_2^{\Lambda\bar{\Sigma}^0} &=& \kappa - \alpha_{\rho(3D)}^{\Lambda\bar{\Sigma}^0},
\end{eqnarray}
where $\kappa = -1.98$~\cite{Leinweber:1990dv,Aliev:2001uq} and $(\gamma, \beta_\rho^{\Lambda\bar{\Sigma}^0}, \beta_{\rho(3D)}^{\Lambda\bar{\Sigma}^0}, \alpha_{\rho(3D)}^{\Lambda\bar{\Sigma}^0})$ are free parameters.

\section{Numerical results} \label{sec:numerical results}

\subsection{The EMFFs of $\Sigma$ hyperon}

We perform a $\chi^2$-fit to the experimental data of $e^+e^-\to\Sigma\bar{\Sigma}$ reactions, including the total cross section, the $|G_E/G_M|$ ratio and the relative phase $\Delta \Phi$. There are 43 data points in total of cross section from BESIII~\cite{BESIII:2020uqk,BESIII:2021rkn,BESIII:2023ldb}, Belle~\cite{Belle:2022dvb} and BaBar~\cite{BaBar:2007fsu}, 6 data points of $|G_E/G_M|$ from BESIII~\cite{BESIII:2020uqk,BESIII:2023ynq}, and 4 data points of $\Delta \Phi$ from BESIII Collaboration~\cite{BESIII:2023ynq}. The are total 10 free parameters. The fitted parameters for both cases of Scenario \uppercase\expandafter{\romannumeral1} and Scenario \uppercase\expandafter{\romannumeral2} are presented in Table~\ref{tab:para}, and the obatined $\chi^2/\rm d.o.f$ are 0.9 and 1.1, respectively. 

\begin{table*}[htbp]
    \centering
 \caption{Fitted parameters in this work.}
    \begin{tabular}{cccccccccccc}\hline\hline
    \multirow{2}*{Scenario \uppercase\expandafter{\romannumeral1}} &Parameter& $\gamma$ & $\beta_{\rho}$ & $\beta_{\omega \phi}$ & $\alpha_{\omega \phi}$ & $\beta_{\omega(3D)}$ & $\alpha_{\omega(3D)}$ & $\beta_{\phi(3D)}$ & $\alpha_{\phi(3D)}$ & $\beta_{\rho(6D)}$ & $\alpha_{\rho(6D)}$\\
    & Value &  0.4217 & 0.9354 & -0.3363 & -2.8767 & 0.6881 & -0.7078 &-0.4998 &  0.4856&-0.0265 & 0.0278  \\ \hline
    \multirow{2}*{Scenario \uppercase\expandafter{\romannumeral2}} &Parameter& $\gamma$ & $\beta_{\rho}$ & $\beta_{\omega \phi}$ & $\alpha_{\omega \phi}$ & $\beta_{\rho(3D)}$ & $\alpha_{\rho(3D)}$ & $\beta_{\phi(3D)}$ & $\alpha_{\phi(3D)}$ & $\beta_{\rho(6D)}$ & $\alpha_{\rho(6D)}$\\
    & Value &  0.4662 & 1.4189 & -0.4264 & 7.9466 & 0.0645 & -0.0968 & 0.3754 &-0.3608 & 0.0537 &  -0.050 \\
    \hline\hline
    \end{tabular}
       \label{tab:para}
\end{table*}

\begin{figure}[htbp]
    \centering
    \includegraphics[scale=0.38]{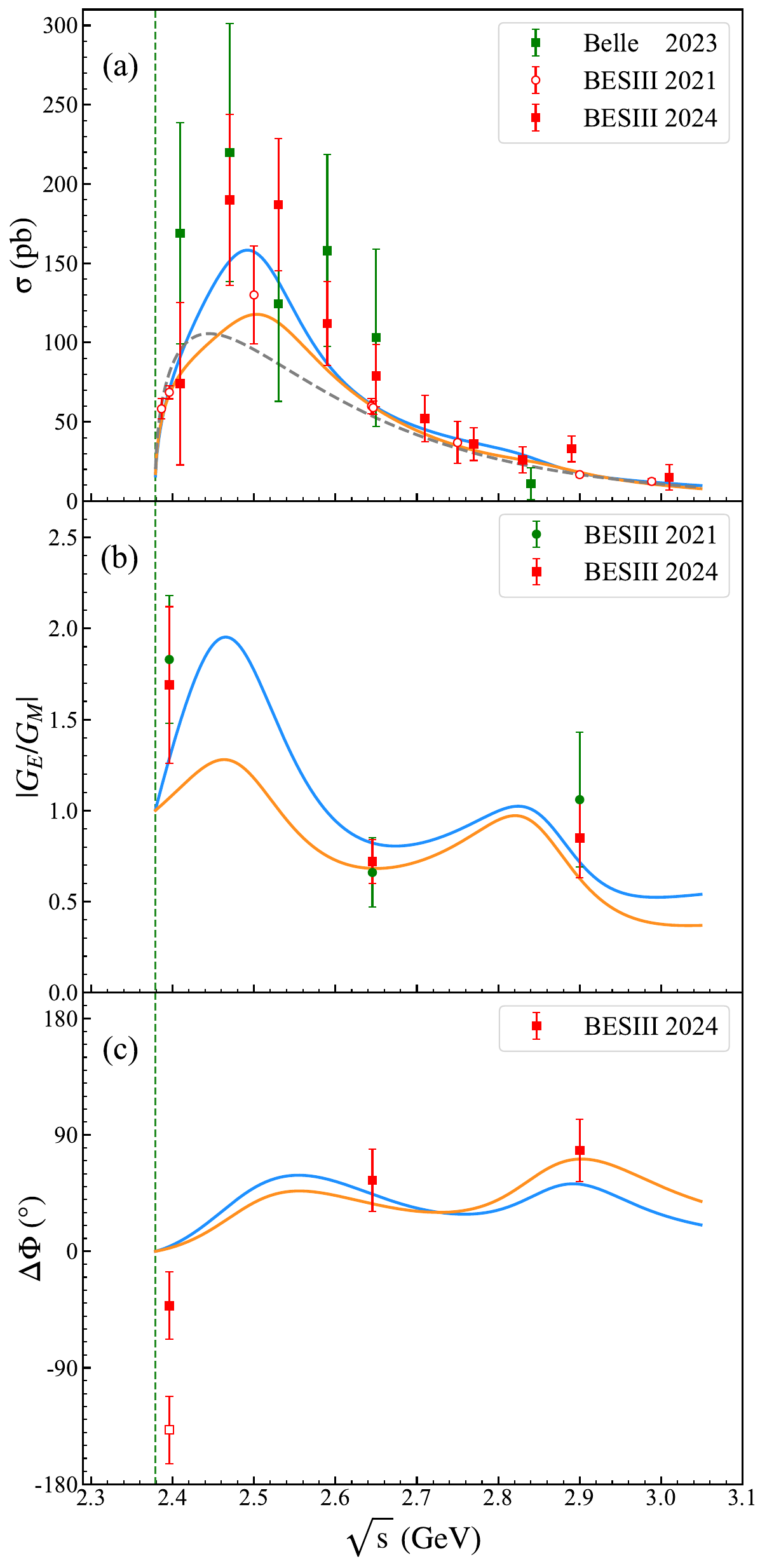}
    \caption{The obtained total cross sections, ratio $|G_E/G_M |$ and relative phase $\Delta \Phi$ of $\Sigma^+$ comparing with the experimental data. The blue solid line stands for the Scenario \uppercase\expandafter{\romannumeral1}, and orange for Scenario \uppercase\expandafter{\romannumeral2}, while the gray dashed line represents our previous work~\cite{Yan:2023yff}. The experimental data are taken from: BESIII 2021~\cite{BESIII:2020uqk}, Belle 2023~\cite{Belle:2022dvb}, BESIII 2024~\cite{BESIII:2023ldb,BESIII:2023ynq}.For the relative phase $\Delta \Phi$ of the red square hollows represented the second solution given by BESIII~\cite{BESIII:2023ynq}.}
    \label{fig:sigmap}
\end{figure}

In Fig.~\ref{fig:sigmap}, we compare our theoretical results with experimental data for the reaction $e^+e^- \to\Sigma^+\bar{\Sigma}^-$. It can be seen that by introducing these vector meson resonance states we obtain a good description of the total cross section as shown in Fig.~\ref{fig:sigmap}(a), the ratio $|G_E/G_M|$ in Fig.~\ref{fig:sigmap}(b) and the relative phase $\Delta \Phi$ in Fig.~\ref{fig:sigmap}(c). Both Scenario \uppercase\expandafter{\romannumeral1} and Scenario \uppercase\expandafter{\romannumeral2} can describe the current experimental data. In the VMD model, considering the two mesons near the reaction threshold, in Fig.~\ref{fig:sigmap}(a) the blue solid line and the orange one show a significant enhancement around the center-of-mass energy $\sqrt{s} = 2.5\ \rm GeV$ compared to the gray line, which only includes the ground states $\rho$, $\omega$, and $\phi$. Notably, the blue aligns well with the latest data in 2024 from BESIII Collaboration in this region. This enhancement can be naturally attributed to the contributions from $\omega(3D)/\rho(3D)$ and $\phi(3D)$. Moreover, we determined the ratio $|G_E/G_M|$ and compared it with the experimental data as illustrated in Fig.~\ref{fig:sigmap}(b). One can clearly see that both cases of the blue and orange lines exhibit the same decreasing oscillatory pattern. This behavior is strikingly similar to the $e^+e^-\to \Lambda_c^+\bar{\Lambda}_c^-$ reaction reported by BESIII~\cite{BESIII:2023rwv}, and the interpretation of the $|G_E/G_M|$ oscillations of $\Lambda_c^+$ is given by the charmonium-like vector mesons near the threshold of $\Lambda_c^+\bar{\Lambda}_c^-$~\cite{Chen:2023oqs}, which here is originating from $\omega(3D)/\rho(3D)$, $\phi(3D)$ and $\rho(6D)$. Remarkably, the blue line agrees more closely with the experimental data near the threshold compared to the orange line, suggesting that the vector particles with isospin 0 may play a dominant role in the $e^+e^-\to \Sigma^+\bar{\Sigma}^-$ process.

In the timelike region, the Breit-Wigner form we use a non-zero width in our model as in Eq.~\ref{eq:bw}, which will lead to a nontrivial phase to the $F_1$ and $F_2$ as well as $G_E$ and $G_M$. Subsequently, we have calculated the relative phase between $G_E$ and $G_M$ depicted in Fig.~\ref{fig:sigmap}(c), and the data comes from the BESIII collaboration. The data at the threshold provided by the BESIII gives two possible results~\cite{BESIII:2023ynq}. Our theoretical results align well with experimental data within the margin of error. Interestingly, the phase observed at 2.9 GeV primarily stems from the interference of the $\rho(6D)$ state with other terms, which would diminish to zero without this state. To match the experimental data, we adopted a relatively large width for the $\rho(6D)$. However, this width significantly exceeds the ranges predicted in Refs.~\cite{Bai:2023dhc,Feng:2022hwq}. Additionally, literature ~\cite{Bai:2023dhc} and ~\cite{Feng:2022hwq} highlight the presence of several excited $\rho$ states in the 2.4–3.0 GeV energy range, which could collectively influence this reaction process. Incorporating these particles into the VMD framework poses challenges, as their parameters cannot be well-constrained due to insufficient data. Therefore, a larger width for the $\rho(6D)$ was used to effectively account for the combined contributions of these vector particles as a single state. Nonetheless, further theoretical and experimental efforts are necessary to gain a deeper understanding of these highly excited vector states.

\begin{figure}[htbp]
    \centering

    \includegraphics[scale=0.38]{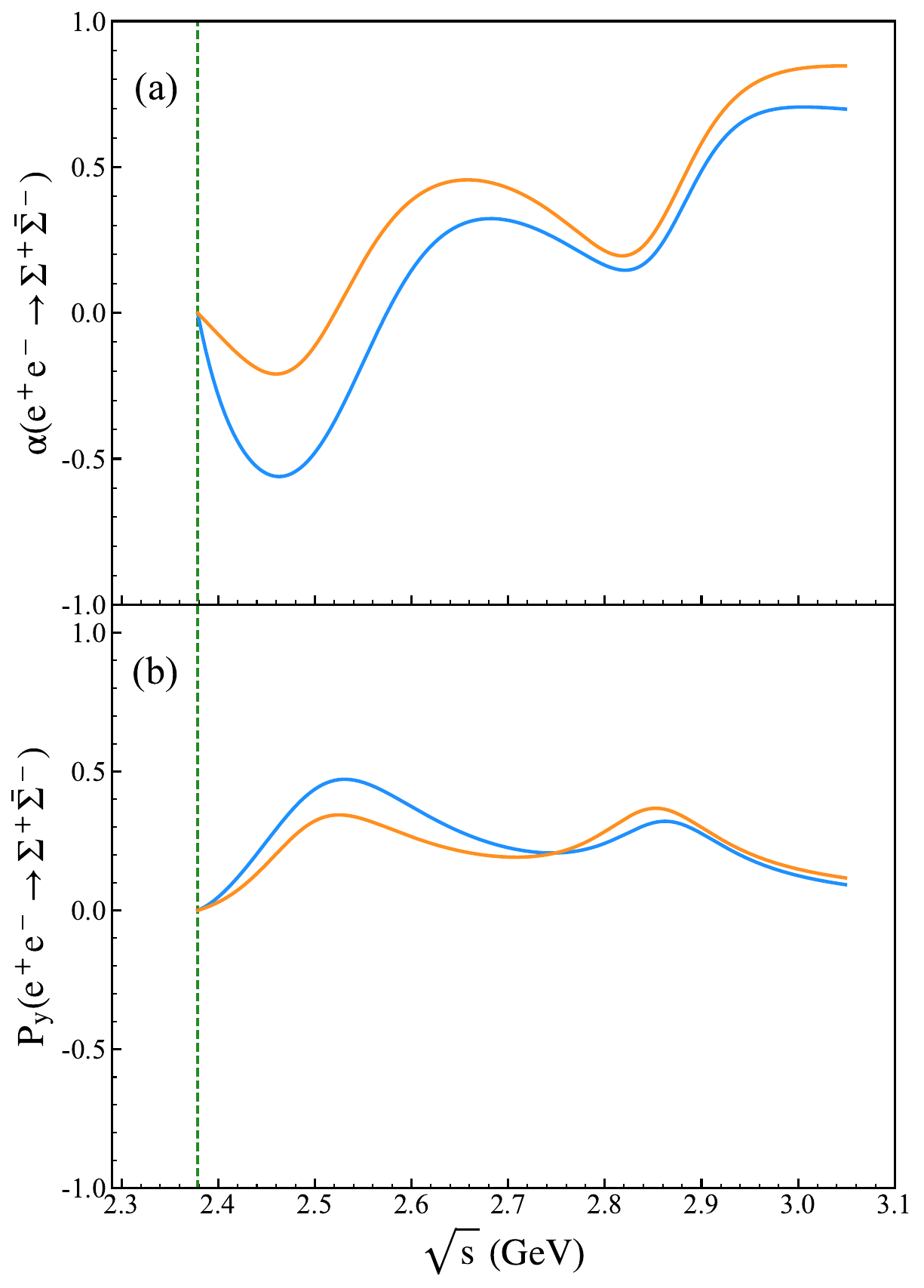}

    \caption{The angular distribution parameter $\alpha$ and polarization $P_y$ of $\Sigma^+$. The blue solid line stands for the Scenario \uppercase\expandafter{\romannumeral1}, and orange for Scenario \uppercase\expandafter{\romannumeral2}.}
    \label{fig:sigmap-alpha-py}
\end{figure}

With the obtained $|G_E/G_M|$, we also calculated the angular distribution parameter $\alpha$ of $\Sigma^+$ in the $e^+ e^- \to \Sigma^+ \bar{\Sigma}^-$ reaction, which are shown in Fig.~\ref{fig:sigmap-alpha-py} (a). In addition, with the obtained relative phase, the spin polarization of $\Sigma^+$ in the timelike region can be estimated with Eq.~\ref{eq:pol}. In Fig.~\ref{fig:sigmap-alpha-py}(b), we present the polarization observable $P_y$ as a function of $\sqrt{s}$, at a fixed scattering angle $\theta = 45^\circ$. One can see that the shape of $P_y$ will be similar to the $\Delta \Phi$ as shown in Fig.~\ref{fig:sigmap}(c) since $P_y$ is proportional to $\sin\Delta \Phi$. Besides, in both cases, the polarization peak appears almost at the resonances $\phi(3D)$ and $\rho(6D)$, which hopefully can be tested by future experiments, especially near the energy region $\sqrt{s} = 2.5 \, \rm GeV$. It is expected that these predictions could be tested by future experiments.

\begin{figure}[htbp]
    \centering
    \includegraphics[scale=0.38]{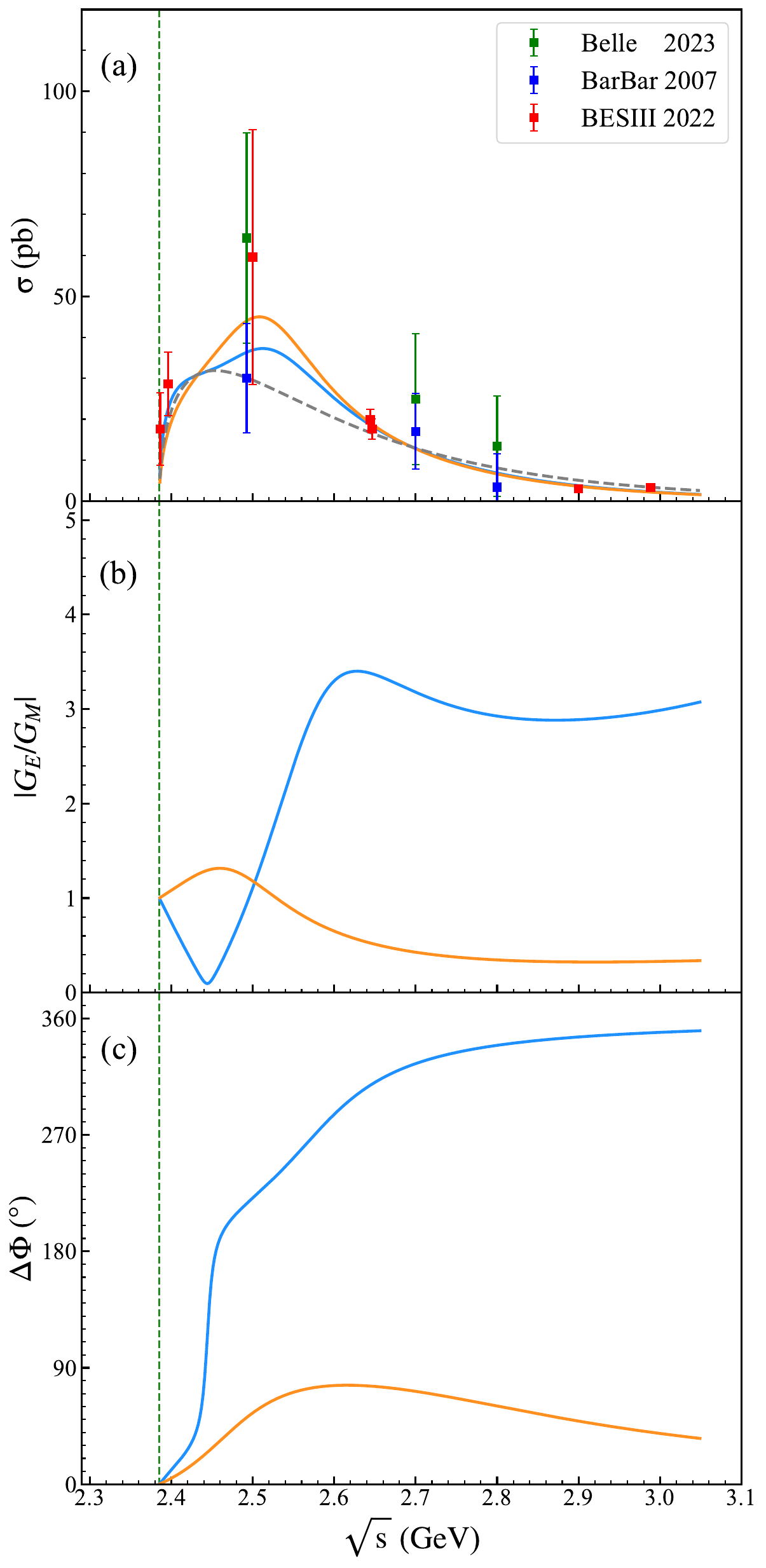}

    \caption{The obtained total cross sections of $\Sigma^0$ comparing with the experimental data, and the prediction for the ratio $|G_E/G_M |$ and relative phase $\Delta \Phi$. The blue solid line stands for the Scenario \uppercase\expandafter{\romannumeral1}, and orange for Scenario \uppercase\expandafter{\romannumeral2}, while the gray dashed line represents our previous work~\cite{Yan:2023yff}. The experimental data are taken from: Belle~\cite{Belle:2022dvb} and BarBar~\cite{BaBar:2007fsu} and  BESIII~\cite{BESIII:2021rkn}.}
    \label{fig:sigma0}
\end{figure}

\begin{figure}[htbp]
    \centering

    \includegraphics[scale=0.38]{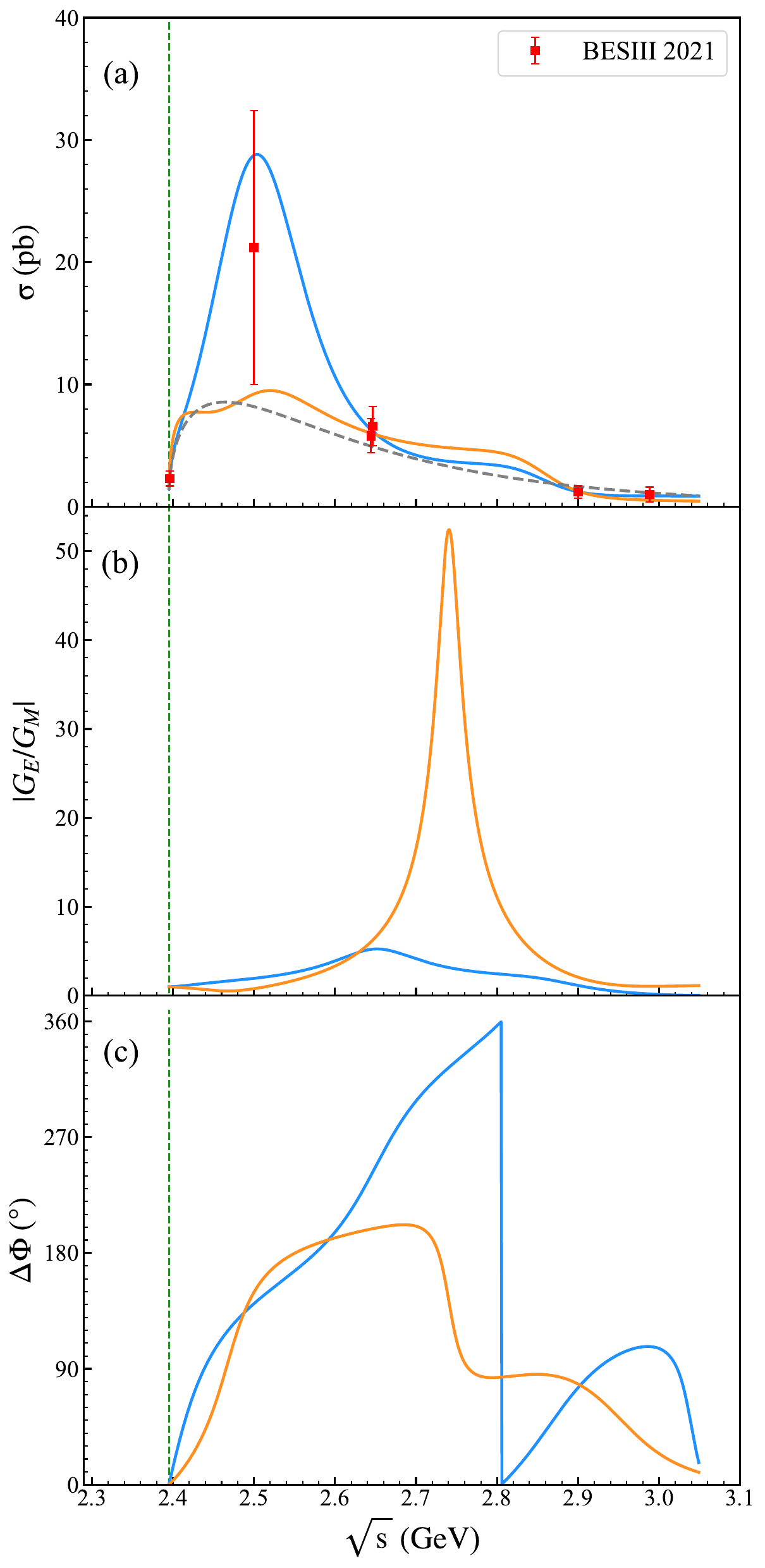}

    \caption{The obtained total cross sections of $\Sigma^-$ comparing with the experimental data, and the prediction for the ratio $|G_E/G_M |$ and relative phase $\Delta \Phi$. The blue solid line stands for the Scenario \uppercase\expandafter{\romannumeral1}, and orange for Scenario \uppercase\expandafter{\romannumeral2}, while the gray dashed line represents our previous work~\cite{Yan:2023yff}. The experimental data are taken from: BESIII 2021~\cite{BESIII:2020uqk}.}
    \label{fig:sigman}
\end{figure}

For the processes $e^+e^-\to \Sigma^0\bar{\Sigma}^0$ and $e^+e^-\to \Sigma^-\bar{\Sigma}^+$, the theoretical results are illustrated in Figs.~\ref{fig:sigma0} and ~\ref{fig:sigman}, respectively. The blue solid line represents Scenario \uppercase\expandafter{\romannumeral1}, while the orange line denotes Scenario \uppercase\expandafter{\romannumeral2}. Currently, only experimental data on their total cross section are available. As shown in Fig.~\ref{fig:sigma0}(a) the cross section of $\Sigma^0$ exhibits a notable enhancement around 2.5 GeV. And comparable enhancement is also seen for $\Sigma^-$, especially for Scenario \uppercase\expandafter{\romannumeral1} (the blue solid line), as depicted in Fig.~\ref{fig:sigman}(a). This enhancement at 2.5 GeV is attributed to the combined contributions of subthreshold $\rho(3D)$, $\omega(3D)$, and $\phi(3D)$ states for the same reason as in the reaction $e^+e^-\to \Sigma^+\bar{\Sigma}^-$. Again, we emphasize that these three vector mesons near the threshold are important to the cross section enhancement in the reactions $e^+e^-\to \Sigma\bar{\Sigma}$. Additionally, it is feasible to consider both isospins near 2.3 GeV, just below the sigma threshold. Furthermore, Ref.~\cite{Bai:2023dhc} predicts that the branching ratio for the decay of $\phi(3D)$ to $\Sigma\bar{\Sigma}$ are larger than those of the higher excited $\rho$ mesons predicted by the MGI model~\cite{Feng:2022hwq}, which further supports our findings. 

Theoretical calculations for the ratio $|G_E/G_M|$ and relative phase $\Delta \Phi$ of $\Sigma^0$ and $\Sigma^-$ are presented. As shown in Figs.~\ref{fig:sigma0}(b) and ~\ref{fig:sigma0}(c) these two quantities for $\Sigma^0$ differ significantly between the two parameterizations (Scenario \uppercase\expandafter{\romannumeral1} and Scenario \uppercase\expandafter{\romannumeral2}), highlighting their sensitivity to the isospin of the vector meson excited states, specifically $\rho(3D)$ and $\omega(3D)$ here. The large discrepancies between the two parameterizations for $\Sigma^0$ and $\Sigma^-$ can be primarily attributed to the isospin decomposition effects. The same parameter set was used to describe the electromagnetic form factors of the $\Sigma$ triplet states, with the contributions of isospin $I=0$ and $I=1$ constrained by the vector meson dominance framework. For $\Sigma^0$, the $I=1$ component does not contribute, while for $\Sigma^-$, the $I=1$ component introduces an additional negative sign relative to the $I=0$ component. Although substantial experimental data on cross sections for $e^+e^-\to \Sigma\bar{\Sigma}$ processes are available, data on the ratios $|G_E/G_M|$ and relative phases $\Delta \Phi$ remain limited. This lack of data imposes relatively weak constraints on the isospin properties of vector mesons near the subthreshold region. Future experimental and theoretical investigations are crucial to clarify the individual contributions of these vector mesons.

For $\Sigma^0$, in Fig.~\ref{fig:sigma0}(b) the blue curve decreases and then rises as the center-of-mass energy increases, whereas the orange curve shows the opposite behavior, rising first and then falling. Beyond 2.7 GeV, both curves flatten out because there is no contribution from $\rho(6D)$ in the $e^+e^-\to \Sigma^0\bar{\Sigma}^0$ process. While for $\Sigma^-$, the obtained ratio $|G_E/G_M|$ and the relative phase $\Delta \Phi$ are shown in Fig.~\ref{fig:sigman}(b) and \ref{fig:sigman}(c), respectively. In Fig.~\ref{fig:sigman}(b), the orange solid line exhibits a pronounced peak in the 2.7–2.8 GeV range, indicating that the angular distribution parameter approaches -1 in this region.

\begin{figure}[htbp]
    \centering

    \includegraphics[scale=0.38]{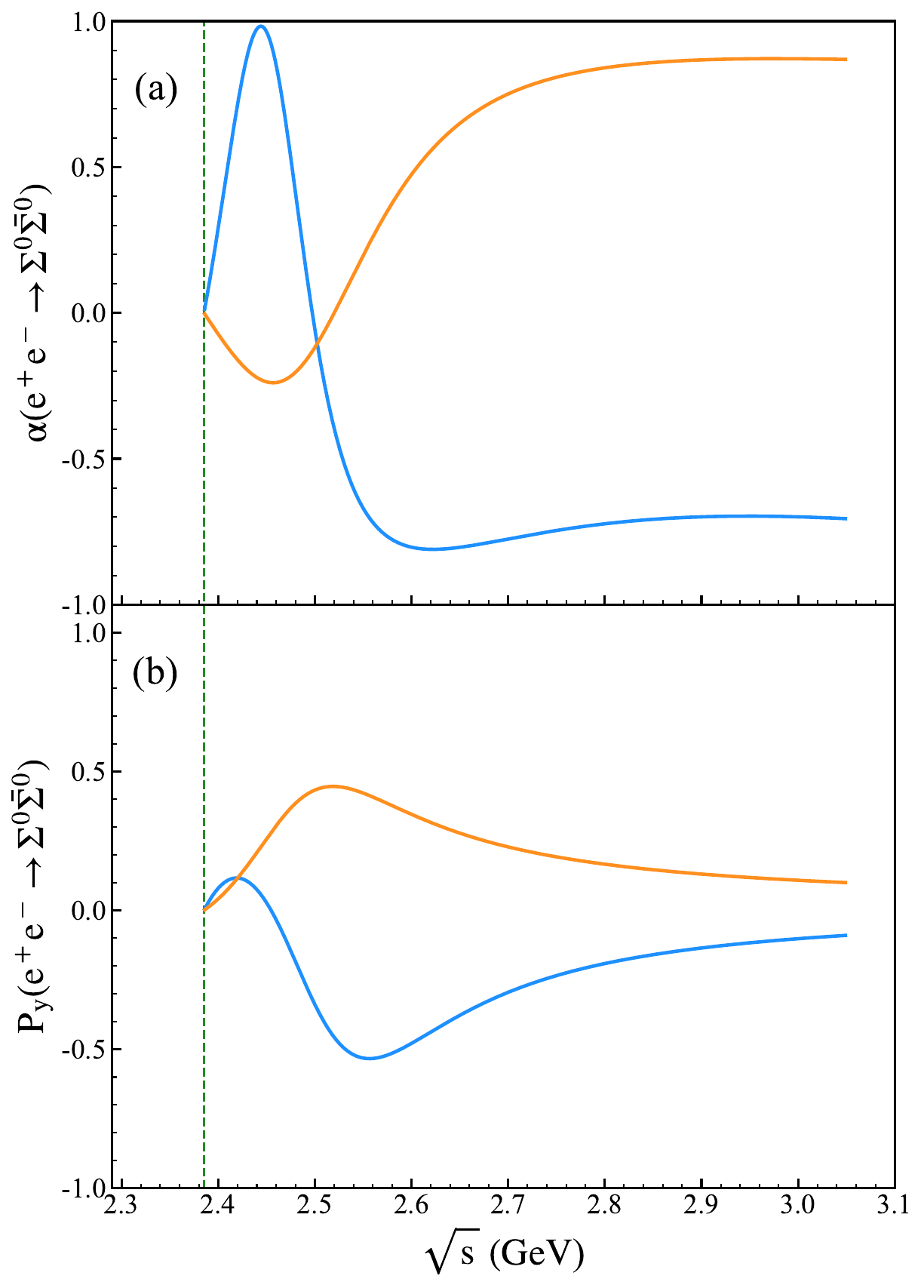}

    \caption{The angular distribution parameter $\alpha$ and polarization $P_y$ of $\Sigma^0$.The blue solid line stands for the Scenario \uppercase\expandafter{\romannumeral1}, and orange for Scenario \uppercase\expandafter{\romannumeral2}.}
    \label{fig:sigma0-alpha-py}
\end{figure}

\begin{figure}[htbp]
    \centering

    \includegraphics[scale=0.38]{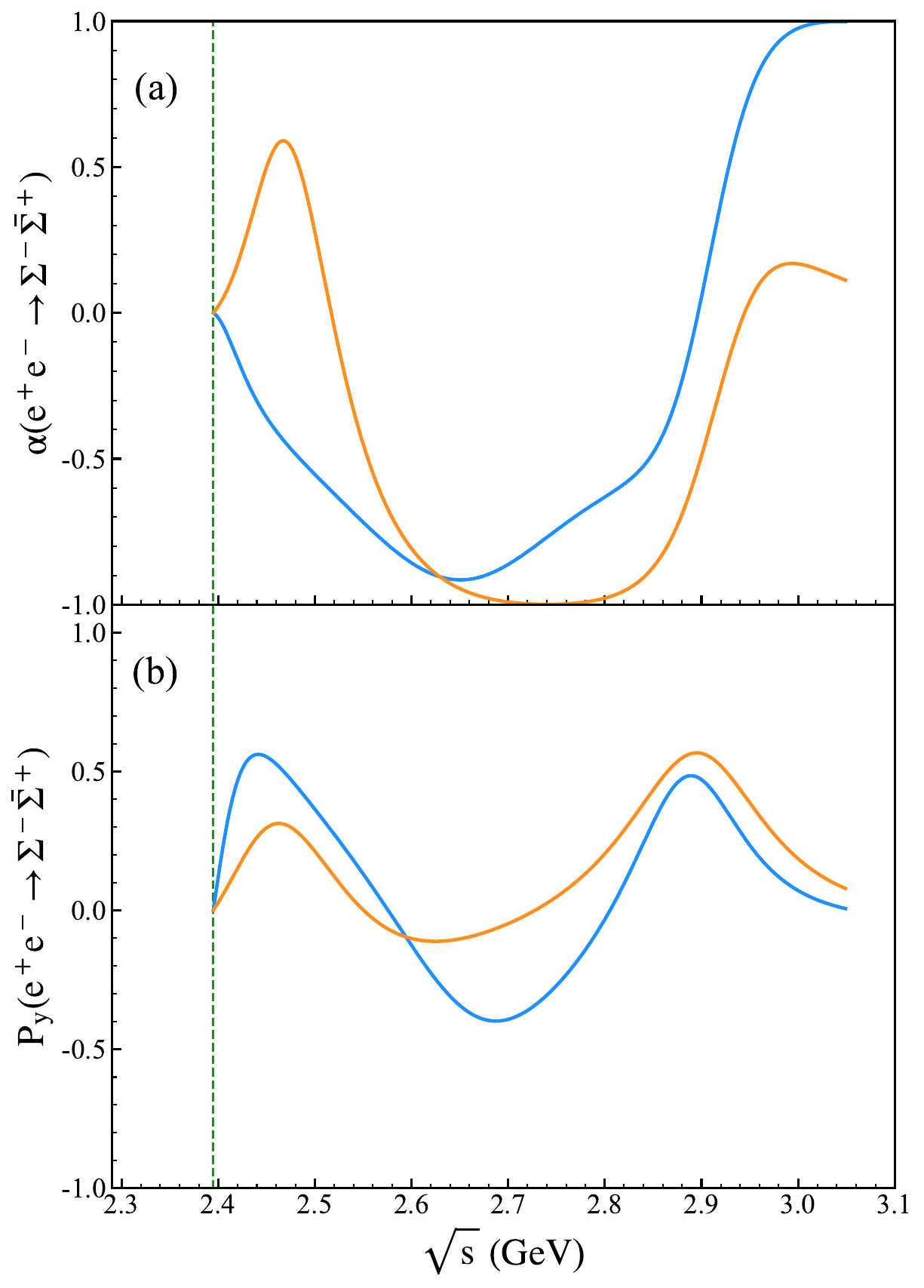}

    \caption{The angular distribution parameter $\alpha$ and polarization $P_y$ of $\Sigma^-$.The blue solid line stands for the Scenario \uppercase\expandafter{\romannumeral1}, and orange for Scenario \uppercase\expandafter{\romannumeral2}.}
    \label{fig:sigman-alpha-py}
\end{figure}

As well, we get the results of $\alpha$ and $p_y$ for $\Sigma^0$ and $\Sigma^-$, as displayed in Figs.~\ref{fig:sigma0-alpha-py} and \ref{fig:sigman-alpha-py}. It can be seen that the differential cross section for Scenario \uppercase\expandafter{\romannumeral1} and Scenario \uppercase\expandafter{\romannumeral2} exhibit different angular distributions. Therefore, measuring the differential cross section of the reactions $e^+e^-\to \Sigma^0\bar{\Sigma}^0$ and $e^+e^-\to \Sigma^- \bar{\Sigma}^+$ are very useful for determining the different parameterizations. Note that in Fig.~\ref{fig:sigma0}(c), the phase $\Delta \Phi$ in Scenario \uppercase\expandafter{\romannumeral1} changes rapidly by $180^\circ$ within 80 MeV above the threshold, whereas in Scenario \uppercase\expandafter{\romannumeral2} it varies more smoothly. After 2.7 GeV, both cases exhibit relatively stable phases. The corresponding polarization $P_y$ of $\Sigma^0$ is shown in Fig.~\ref{fig:sigma0-alpha-py}(b). Near the threshold, both cases possess the same polarization behavior but show completely opposite results after 2.45 GeV and will present significant polarization effects at 2.5-2.6 GeV. On the other hand, from Fig.~\ref{fig:sigman-alpha-py}(b), it is evident that the peak polarization of $\Sigma^-$ occurs near the $\phi(3D)$ and $\rho(6D)$ resonance states, with a variation pattern closely resembling the polarization evolution of $\Sigma^+$.

\subsection{The $\Lambda \to \bar{\Sigma}^0$ transition EMFFs}

In the $e^+e^- \to \Lambda \bar{\Sigma}^0$ reaction, the total isospin of the $\Lambda$ and $\bar{\Sigma}^0$ system is 1, which means that only vector particles with isospin 1 contribute to the process in the VMD model. Thus, the $\rho$ and excited $\rho(3D)$ state are introduced, resulting in four parameters $(\gamma, \beta_{\rho}^{\Lambda\bar{\Sigma}^0}, \beta_{\rho(3D)}^{\Lambda\bar{\Sigma}^0}, \alpha_{\rho(3D)}^{\Lambda\bar{\Sigma}^0})$ that need to be determined. Firstly, we fixed $\gamma$ at 0.43 for the intrinsic form factor $g(q^2)$. This form factor reflects the internal structure of the three valence quarks within the baryon~\cite{Bijker:1995ii}. According to the quark model, both $\Sigma$ and $\Lambda$ particles contain one strange quark and two light-flavor quarks, with the same flavor structure. Furthermore, previous studies on the $e^+e^- \to \Lambda \bar{\Lambda}$ reaction report a $\gamma$ value of 0.43~\cite{Li:2021lvs}, which aligns well with the $\gamma$ value for $\Sigma$ from our work (between 0.4217 and 0.4662 in different scenarios). Given these similarities, we assumed the $\gamma$ parameter for the $e^+e^- \to \Lambda \bar{\Sigma}^0$ process were consistent with values from reactions $e^+e^- \to \Sigma \bar{\Sigma}$ and $e^+e^- \to \Lambda \bar{\Lambda}$, and hence fixed at 0.43. Besides, in Ref.~\cite{PhysRev.155.1562}, it was found that the vector coupling constant of $g_{\Sigma\Lambda\rho}$ to be 0 employing the method of current algebra. The same result was obtained using the QCD sum rule~\cite{Erkol:2006sa}. Since $\beta_{\rho}^{\Lambda\bar{\Sigma}^0}$ is proportional to $g_{\Sigma\Lambda\rho}$, we fixed the parameter $\beta_{\rho}^{\Lambda\bar{\Sigma}^0}=0$. For the remaining parameters, we determined by fitting them to the total cross-section data of the $e^+e^-\to \Lambda\bar{\Sigma}^0$ reaction from BESIII Collaboration~\cite{BESIII:2023pfv}, excluding BaBar data due to large experimental uncertainties~\cite{BaBar:2007fsu}. The resulting fitted parameters were $\beta_{\rho(3D)}^{\Lambda\bar{\Sigma}^0} = 0.3784$, $\alpha_{\rho(3D)}^{\Lambda\bar{\Sigma}^0} = -0.4053$, with $\chi^2/\rm d.o.f$ achieving a value of 1.0.

\begin{figure}[htbp]
    \centering
    \includegraphics[scale=0.38]{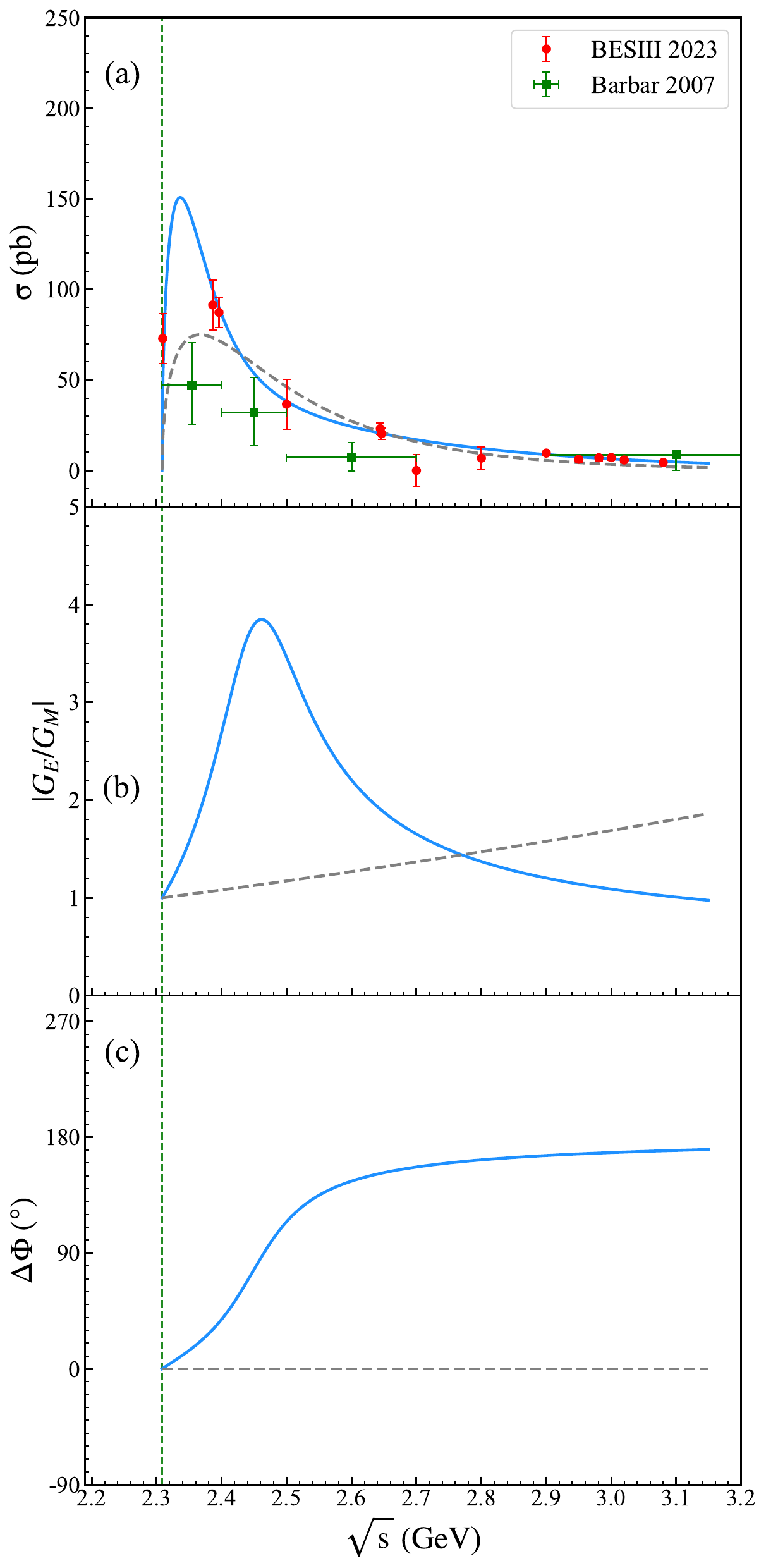}

    \caption{The obtained total cross sections of the process $e^+e^-\to \Lambda\bar{\Sigma}^0$ comparing with the experimental data, and the prediction for the ratio $|G_E/G_M |$ and relative phase $\Delta \Phi$. The blue solid line represents the inclusion of the $\rho(3D)$, while the dashed line is the exclusion of $\rho(3D)$.The experimental data are taken from: BarBar~\cite{BaBar:2007fsu} and  BESIII~\cite{BESIII:2023pfv}.}
    \label{fig:sila}
\end{figure}

In Fig.~\ref{fig:sila}, we compare our fit to the experimental data. The blue solid line includes the contribution of $\rho(3D)$, while the gray dashed line represents a fit without $\rho(3D)$ (achieved by setting the $\beta_{\rho(3D)}$ and $\alpha_{\rho(3D)}$ coupling constant to zero). As shown in Fig.~\ref{fig:sila}(a), we can see that the blue solid line and the gray dashed line provide a reasonable description of the total cross-section. However, the blue solid line shows a notable enhancement near the threshold, with cross-sections reaching approximately 150 pb. This increase is attributed to the influence of $\rho(3D)$ near the threshold, consistent with certain theoretical explanations for the threshold enhancement in the $e^+e^- \to \Lambda \bar{\Lambda}$ process~\cite{Li:2021lvs,Cao:2018kos,Xiao:2019qhl}

\begin{figure}[htbp]
    \centering
    \includegraphics[scale=0.38]{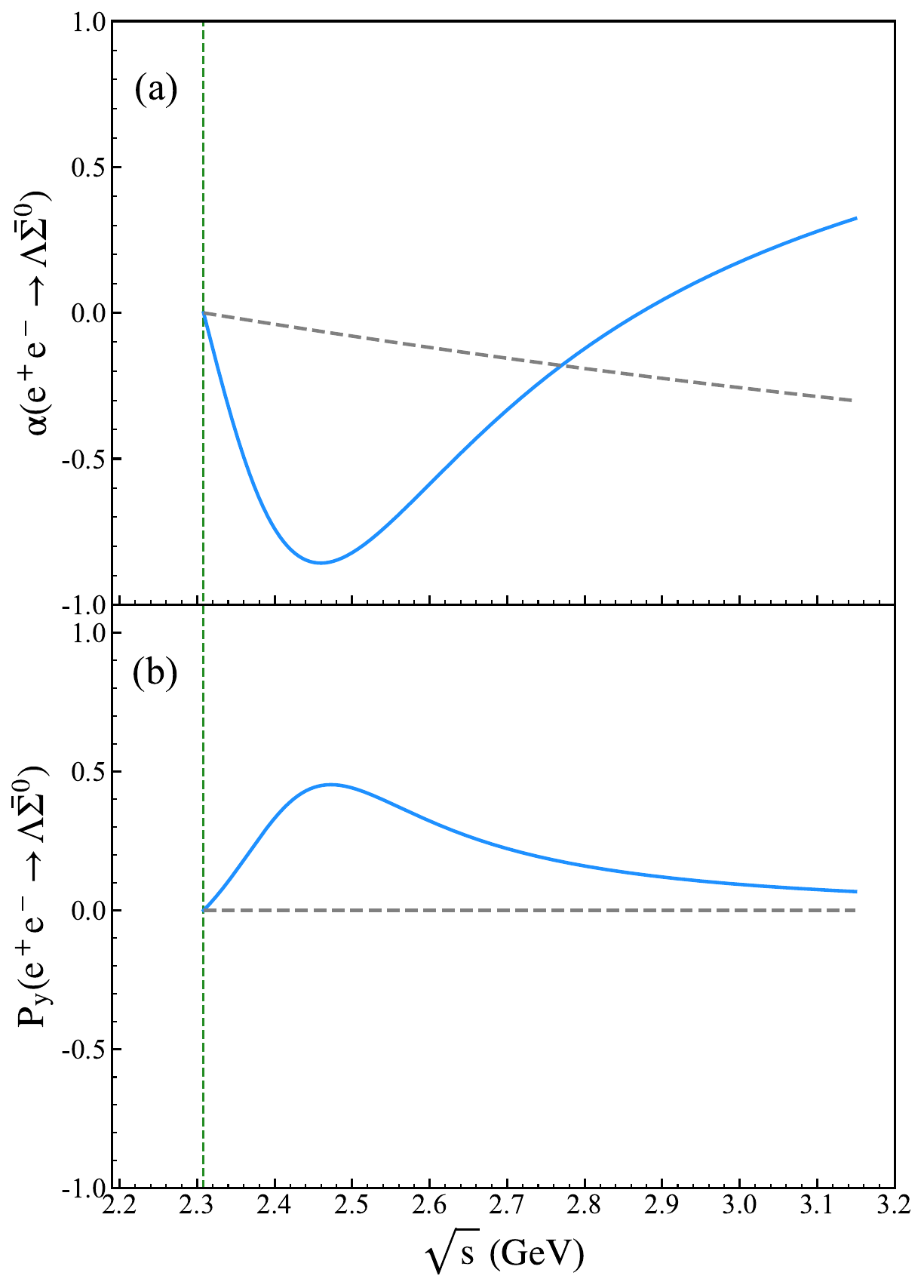}

    \caption{The obtained angular distribution parameter $\alpha$ and polarization $P_y$ of $\Lambda\bar{\Sigma}^0$.The blue solid line represents the inclusion of the $\rho(3D)$, while the dashed line is the exclusion of $\rho(3D)$.}
    \label{fig:sila-alpha-py}
\end{figure}

As a prediction, we also present the electromagnetic ratio $|G_E/G_M|$ and relative phase $\Delta \Phi$ for the $\Lambda-\bar{\Sigma}^0$ system, shown in Figs.~\ref{fig:sila}(b) and ~\ref{fig:sila}(c), respectively. In Fig.~\ref{fig:sila}(b), the blue solid line, which includes $\rho(3D)$, exhibits a distinct non-monotonic behavior compared to the gray dashed line without $\rho(3D)$. Starting near the threshold, the ratio $|G_E/G_M|$ first increases, then decreases, eventually stabilizing at 1 in the higher energy region. One can note that the ratio $|G_E/G_M|$ approaches large value at a center-of-mass energy of $\sqrt{s} = 2.46 \, \mathrm{GeV}$, where the angular distribution parameter $\alpha$ converges to -1, as shown in Fig.~\ref{fig:sila-alpha-py}(a). This indicates the possibility of observing a clear angular distribution for the final state in the reaction $e^+e^- \to \Lambda \bar{\Sigma}^0$ near this energy. Additionally, as shown in Fig.~\ref{fig:sila}(c), the relative phase increases steadily from $0^\circ$ to $180^\circ$ as the center-of-mass energy rises, reaching $90^\circ$ at 2.46 GeV and suggesting the possibility of significant polarization effects. Finally, in Fig.~\ref{fig:sila-alpha-py}(b) we give the results of the polarization $P_y$ exhibits a pronounced peak within the energy range of 2.4-2.5 GeV. It is expected that more precise experimental measurements in the near future can be used to check our findings.

\section{SUMMARY}
In this work, we study the electromagnetic form factors of $\Sigma$ and $\Lambda-\bar{\Sigma}^0$ transition form factors in the timelike region using a phenomenological extending vector meson dominance model. In the reaction $e^+e^-\to\Sigma\bar{\Sigma}$, in addition to ground mesons $\rho$, $\omega$ and $\phi$, we take account into additional vector meson excited states $\rho(3D)$, $\omega(3D)$, $\phi(3D)$ and $\rho(6D)$. We split into two scenarios according to the contribution of $\rho(3D)$ and $\omega(3D)$. With these vector mesons, both scenarios can successfully describe the total cross section of the $\Sigma$ isospin triplet state, as well as the electromagnetic ratio $|G_E/G_M|$ and relative phase $\Delta \Phi$ of $\Sigma^+$. Our results show a significant enhancement of the total cross section around the center-of-mass energy of 2.5 GeV compared to only the ground state vector mesons, which is mainly due to the effect of $\phi(3D)$. This also gives a natural explanation for the cross section recently measured by BESIII. Furthermore, we find that the electromagnetic ratio of $\Sigma^+$ exhibits a decreasing oscillatory behavior, a feature very similar to that in $\Lambda_c^+$. This non-monotonic behavior is very relevant to the vector meson resonance above the reaction threshold. 

In the process of $e^+e^-\to\Lambda\bar{\Sigma}^0$, we have considered the contributions from $\rho$ and $\rho(3D)$ states. Our analysis finds that the $\rho(3D)$ state plays a crucial role in the threshold enhancement of the total cross section. Notably, despite the $\rho(3D)$ lying below the reaction threshold, it significantly impacts the electromagnetic form factors above the threshold. This effect is particularly pronounced in the ratio $|G_E/G_M|$ and its associated $\Delta \Phi$ across a wide energy range.

Finally, we emphasize that in the $e^+e^- \to \Sigma\bar{\Sigma}$ and $\Lambda\bar{\Sigma}^0$ process, its cross section enhancement, the non-monotonic structure of the ratio $|G_E/G_M|$ and the non-zero relative phase $\Delta \Phi$ can be naturally explained by the vector meson dominance model. In particular, the contribution of vector meson resonance states above the reaction threshold is important.

\begin{acknowledgments}

We would like to thank Prof. Xiong-Fei Wang for fruitful discussions. This work is partly supported by the National Key R\&D Program of China under Grant No. 2023YFA1606703, and by the National Natural Science Foundation of China under Grant Nos. 12435007, 12361141819, 12275185 and 12335002.

\end{acknowledgments}


\setcounter{equation}{0}
\renewcommand{\theequation}{A\arabic{equation}}
\setcounter{figure}{0}
\renewcommand{\thefigure}{A\arabic{figure}}

\section*{APPENDIX}
In this appendix,  we present the details of the polarization observables in the reaction $e^+e^- \to \Sigma \bar{\Sigma}$.  The main results about the polarization observables can be found in Refs.~\cite{Czyz:2007wi,Faldt:2013gka,Faldt:2016qee,Faldt:2017kgy}.  To address some notation ambiguities, we provide a detailed derivation of the relevant formulas.

The spin direction of the final-state $\Sigma$ hyperon cannot be directly measured. Instead, it is reconstructed through the decay products of the $\Sigma$. The dominant decay mode for the $\Sigma^+$ and $\Sigma^-$ is into a nucleon and a $\pi$ meson, whereas the $\Sigma^0$ decays into a $\Lambda$ hyperon and a photon. Without loss of generality, let us consider the specific reaction $e^+e^- \to \Sigma^+ (\to p \pi^0) \bar{\Sigma}^- (\to \bar{p} \pi^0)$ in the center-of-mass frame of the $e^+e^-$ system. The total amplitude for this process is given by

\begin{eqnarray}
    \mathcal{M} &= \frac{e^2}{q^2}D_1D_2\bar{v}(k_2)\gamma^\mu u(k_1) H_\mu,
\end{eqnarray}
where $q = k_1+k_2$. Here, $k_1$ and $k_2$ represent the four-momenta of the electron and positron, respectively. Additionally, the propagators are expressed as
\begin{align}
    D_1 = \frac{1}{p_1^2 - M^2 + i M\Gamma},\\
    D_2 = \frac{1}{p_2^2 - M^2 + i M\Gamma},
\end{align}
$M$ and $\Gamma$ denote the mass and total decay width for $\Sigma^+$ hyperon, while $p_1$ and $p_2$ represent the four-momenta of the $\Sigma^+$ and $\bar{\Sigma}^-$, respectively. The final term $H_\mu$ is expressed as
\begin{eqnarray}
    H_\mu &=& \bar{u}(l_1)(a+b\gamma^5)\times (\slashed{p}_1+M) \Gamma_\mu \nonumber \\
    &&\times (\slashed{p}_2-M)(\bar{a}+\bar{b}\gamma^5)v(l_2),
\end{eqnarray}
where $l_1$ and $l_2$ represent the four-momentum of the proton and antiproton, respectively. The coupling constants $a$ and $b$ correspond to the $\Sigma^+ \to p + \pi^0$ transition, while $\bar{a}$ and $\bar{b}$ correspond to the $\bar{\Sigma}^- \to \bar{p} + \pi^0$ transition. The vertex describing the interaction between the photon and the $\Sigma^+$ hyperon is given by
\begin{eqnarray}
    \Gamma^\mu = \gamma^\mu F_1(q^2) + i\frac{F_2(q^2)}{2M}\sigma^{\mu\nu}q_\nu,
\end{eqnarray}
where $F_1$ and $F_2$ are the Dirac and Pauli form factors, respectively.

The differential cross section for this process is expressed as
\begin{eqnarray}
    \mathrm{d}\sigma = \frac{1}{4}\frac{1}{2s}|\mathcal{M}|^2  \mathrm{d}\Phi_4,
\end{eqnarray}
where $s = q^2$, and $\mathrm{d}\Phi_4$ denotes the four-body phase-space element given by
\begin{eqnarray}
    \mathrm{d}\Phi_4 &=& (2\pi)^4\delta^4(k_1 + k_2 - l_1-q_1-l_2-q_2)  \nonumber \\
    &&\times\frac{\mathrm{d}^3l_1}{(2\pi)^3 2l_1^0}\frac{\mathrm{d}^3q_1}{(2\pi)^3 2q_1^0}\frac{\mathrm{d}^3l_{2}}{(2\pi)^3 2l_{2}^0}\frac{\mathrm{d}^3q_{2}}{(2\pi)^3 2q_{2}^0}  \nonumber \\
    &=&\frac{1}{2\pi}\mathrm{d}s_1 \frac{1}{2\pi}\mathrm{d}s_2\frac{p}{16\pi^2\sqrt{s}}\mathrm{d} \Omega   \nonumber \\
    &&\times \frac{|\vec{l}_1^\prime|}{16\pi^2\sqrt{s_1}}\mathrm{d} \Omega_p \frac{|\vec{l}_2^\prime|}{16\pi^2\sqrt{s_2}}\mathrm{d} \Omega_{\bar{p}},
\end{eqnarray}
Here, $s_1 = p_1^2 = (l_1 + q_1)^2$ and $s_2 = p_2^2 = (l_2 + q_2)^2$ represent the invariant masses squared of the $\Sigma^+$ and $\bar{\Sigma}^-$ systems, respectively. The variables $q_1$ and $q_2$ correspond to the momentum of the emitted pions. The quantities $(\Omega_p, |\vec{l}_1'|)$ and $(\Omega_{\bar{p}}, |\vec{l}_2'|)$ denote the solid angles and the magnitudes of the momentum of the proton and antiproton in the rest frames of the $\Sigma^+$ and $\bar{\Sigma}^-$, respectively. Additionally, $p = |\vec{p}|$ represents the magnitude of the $\Sigma^+$ momentum, and $\Omega$ corresponds to its solid angle.

For the subsequent analysis, we define the coordinate system and provide the explicit expressions for the four-momenta of the particles. The coordinate axes are defined as:
\begin{eqnarray}
    \mathbf{e}_x = \frac{1}{\sin \theta} (\hat{k} - \hat{p}\cos \theta),\ \ \mathbf{e}_y = \frac{1}{\sin \theta} \hat{p}\times \hat{k},\ \ \mathbf{e}_z = \hat{p},
\end{eqnarray}
where $\hat{k}$ and $\hat{p}$ are unit vectors representing the directions of the electron and $\Sigma^+$ momentum, respectively. The angle $\theta$ between $\hat{k}$ and $\hat{p}$ satisfies $\cos \theta = \hat{k}\cdot\hat{p}$, and we denote the solid angle as $\Omega=(\theta,\phi)$. The four-momenta of the electron and positron are given by:
\begin{eqnarray}
    k_1 &=& (E,E\sin\theta,0,E\cos\theta), \\
    k_2 &=& (E,-E\sin\theta,0,-E\cos\theta), 
\end{eqnarray}
where $E$ is the energy of the electron. The four-momenta of the $\Sigma^+$ and $\bar{\Sigma}^-$ are:
\begin{eqnarray}
    p_1=(E,0,0,p), \qquad p_2=(E,0,0,-p).
\end{eqnarray}

Additionally, we define the following quantities:
\begin{eqnarray}
    P = q, \qquad Q = p_1 - p_2.
\end{eqnarray}
The proton momentum in the $\Sigma^+$ rest frame is expressed as:
\begin{eqnarray}
    \vec{l}_1^\prime&= |\vec{l}_1^\prime|(\sin \theta_p \cos\phi_p, \sin \theta_p \sin \phi_p, \cos \theta_p),
\end{eqnarray}
and the antiproton momentum in the $\bar{\Sigma}^-$ rest frame is:
\begin{eqnarray}
    \vec{l}_2^\prime&= |\vec{l}_2^\prime|(\sin \theta_{\bar{p}} \cos\phi_{\bar{p}}, \sin \theta_{\bar{p}} \sin \phi_{\bar{p}}, \cos \theta_{\bar{p}}),
\end{eqnarray}
where $\Omega_p=(\theta_p,\phi_p)$ and $\Omega_{\bar{p}}=(\theta_{\bar{p}},\phi_{\bar{p}})$ represent the solid angles.

Our primary objective is to calculate $|\mathcal{M}|^2$. The leptonic tensor is defined as:
\begin{eqnarray}
    L^{\mu\nu} &=&\frac{1}{4}\sum_{\rm spins} \bar{v}(k_2)\gamma^\mu u(k_1)\bar{u}(k_1)\gamma^\nu v(k_2) \nonumber  \\
    &=& k_1^\mu k_2^\nu + k_1^\nu k_2^\mu -\frac{1}{2}s g^{\mu\nu}.
\end{eqnarray}
In this definition, $\frac{1}{4}\sum_{\rm spins}$ signifies the summation and averaging over the initial electron and positron spins. Additionally, the lepton masses are neglected in this calculation.

The hadronic tensor is defined as
\begin{eqnarray}
    && H_{\mu\nu} = \sum_{\rm spins}  H_\mu H_\nu^\dagger = \mathrm{Tr}\left[(\slashed{p}_1+M)(a^*-b^*\gamma^5)(\slashed{l}_1 + m)\right. \nonumber \\
    &&\qquad (a+b\gamma^5)(\slashed{p}_1+M)\Gamma_\mu  (\slashed{p}_2-M)(\bar{a}+\bar{b}\gamma^5)(\slashed{l}_2-m) \nonumber \\
     &&\qquad \left.(\bar{a}^*-\bar{b}^*\gamma^5)(\slashed{p}_2 - M)\bar{\Gamma}_\nu \right],
\end{eqnarray}
where the summation $\sum_{\rm spins}$ represents the spin sum over all final protons and antiprotons, and $m$ denotes the proton mass. The quantity $\bar{\Gamma}_\nu$ is defined as
\begin{eqnarray}
    \bar{\Gamma}_\nu = \gamma^0 \Gamma_\nu^\dagger\gamma^0.
\end{eqnarray}

We can simplify the hadronic tensor $H^{\mu\nu}$ by introducing the following definitions. Let
\begin{eqnarray}
    X &=& (\slashed{p}_1+M)(a^*-b^*\gamma^5)(\slashed{l}_1 + m)(a+b\gamma^5)(\slashed{p}_1+M) \nonumber\\
      &=& (\slashed{p}_1+M)\left[ 2p_1\cdot l_1 (|a|^2+|b|^2) + 2Mm(|a|^2-|b|^2) \right.\nonumber\\
      &&\left.- 2(ab^*+a^*b)\gamma^5(M\slashed{l}_1+p_1\cdot l_1)\right]\nonumber\\
      &=& R(\slashed{p}_1+M)\left[ 1 + S \gamma^5 \left( \slashed{l}_1+\frac{p_1 \cdot l_1}{M} \right)\right],
\end{eqnarray}
and
\begin{eqnarray}
    Y &=& (\slashed{p}_2-M)(\bar{a}+\bar{b}\gamma^5)(\slashed{l}_2-m)(\bar{a}^*-\bar{b}^*\gamma^5)(\slashed{p}_2 - M)\nonumber\\
      &=& (\slashed{p}_2-M)\left[ 2p_2\cdot l_2 (|\bar{a}|^2+|\bar{b}|^2) + 2Mm(|\bar{a}|^2-|\bar{b}|^2)\right. \nonumber\\
      &&\left.- 2(\bar{a}\bar{b}^*+\bar{a}^*\bar{b})\gamma^5(M\slashed{l}_2-p_2\cdot l_2)\right]\nonumber\\
      &=& \bar{R}(\slashed{p}_2-M)\left[ 1 + \bar{S} \gamma^5 \left( \slashed{l}_2-\frac{p_2 \cdot l_2}{M} \right)\right],
\end{eqnarray}
where
\begin{eqnarray}
    R &=& 2p_1\cdot l_1 (|a|^2+|b|^2) + 2Mm(|a|^2-|b|^2) \nonumber\\
    &=& |a|^2[(M+m)^2-\mu^2] + |b|^2[(M-m)^2-\mu^2], \qquad \\
    \bar{R} &=& 2p_2\cdot l_2 (|\bar{a}|^2+|\bar{b}|^2) + 2Mm(|\bar{a}|^2-|\bar{b}|^2) \nonumber\\
    &=& |\bar{a}|^2[(M+m)^2-\mu^2] + |\bar{b}|^2[(M-m)^2-\mu^2],
\end{eqnarray}
with $\mu$ the mass of $\pi_0$, and
\begin{eqnarray}
    S &=& - \frac{2M(ab^*+a^*b)}{R},  \\
    \bar{S} &=& - \frac{2M(\bar{a}\bar{b}^*+\bar{a}^*\bar{b})}{\bar{R}}.
\end{eqnarray}
It is evident that the quantities $\frac{p_1 \cdot l_1}{M}$ and $\frac{p_2 \cdot l_2}{M}$ represent the energies of the proton and antiproton in the rest frame of $\Sigma^+$ and $\bar{\Sigma}^-$, respectively. For simplicity, we introduce the following notation:
\begin{eqnarray}
    H_1 = \frac{p_1 \cdot l_1}{M},\qquad H_2 = \frac{p_2 \cdot l_2}{M}.
\end{eqnarray}
Furthermore, the parameters $S$ and $\bar{S}$ are connected to the decay asymmetry parameters of $\Sigma^+ \to p + \pi^0$ and $\bar{\Sigma}^- \to \bar{p} + \pi^0$, which are given by:
\begin{eqnarray}
    \alpha_{\Sigma^+} = S|\vec{l}_1^\prime|, \qquad  \alpha_{\bar{\Sigma}^-} = \bar{S}|\vec{l}_2^\prime|.
\end{eqnarray}

Next, we decompose the hadronic tensor $H^{\mu\nu}$ into four independent components:

\begin{eqnarray}
    H^{\mu\nu} = R\bar{R}\left[ H^{\mu\nu}_{I} + SH^{\mu\nu}_{S} + \bar{S}H^{\mu\nu}_{\bar{S}} + S\bar{S}H^{\mu\nu}_{S\bar{S}} \right].
\end{eqnarray}
The first component is given by
\begin{eqnarray}
     H^{\mu\nu}_{I} &=& \mathrm{Tr}\left[(\slashed{p}_1+M) \Gamma^\mu (\slashed{p}_2-M)\bar{\Gamma}^\nu \right] \nonumber\\
     &=&4|G_M|^2\left(p_1^\mu p_2^\nu + p_1^\nu p_2^\mu - \frac{1}{2}s g^{\mu \nu} \right)\nonumber\\
     &&+2(G_M F_2^*+G_M^*F_2)(p_1^\mu-p_2^\mu)(p_1^\nu-p_2^\nu) \nonumber \\
     && + 2\left(\frac{s}{4M^2}-1\right)|F_2|^2(p_1^\mu-p_2^\mu)(p_1^\nu-p_2^\nu) \nonumber\\
     &=&4|G_M|^2\left(p_1^\mu p_2^\nu + p_1^\nu p_2^\mu - \frac{1}{2}s g^{\mu \nu} \right) \nonumber\\
     &&+ \frac{2}{\tau-1}(|G_E|^2-|G_M|^2)Q^\mu Q^\nu.
\end{eqnarray}
Contracting $H^{\mu\nu}_{I}$ with $L_{\mu \nu}$, we obtain
\begin{eqnarray}
    H^{\mu\nu}_{I}L_{\mu \nu} &=& s^2(1+\cos^2\theta)|G_M|^2 + 4sM^2|G_E^2|\sin^2\theta \nonumber \\
    &=&sD(s)(1+\alpha \cos^2 \theta),
\end{eqnarray}
where
\begin{eqnarray}
    D(s) = s|G_M|^2+4M^2|G_E|^2,   
\end{eqnarray}
and
\begin{eqnarray}
    \alpha = \frac{s|G_M|^2-4M^2|G_E|^2}{s|G_M|^2+4M^2|G_E|^2}.
\end{eqnarray}

The second component is given by
\begin{eqnarray}
    H^{\mu\nu}_{S} &=& \mathrm{Tr}\left[(\slashed{p}_1+M)\gamma^5 \left( \slashed{l}_1+H_1 \right) \Gamma^\mu (\slashed{p}_2-M)\bar{\Gamma}^\nu \right] \nonumber\\
    &=&-4i|G_M|^2(M\varepsilon^{\rho\sigma\mu\nu}({p_1}_{\rho}+{p_2}_{\rho}) {l_1}_\sigma + H_1\varepsilon^{\rho\sigma\mu\nu}{p_1}_{\rho}{p_2}_{\sigma}) \nonumber\\
    &&-G_M\frac{F_2^*}{2M}4i\varepsilon^{\alpha\beta\gamma\mu}{p_1}_\alpha{p_2}_\beta{l_1}_\gamma(p_1^\nu-p_2^\nu) \nonumber\\
    &&+G_M^*\frac{F_2}{2M}4i\varepsilon^{\alpha\beta\gamma\nu}{p_1}_\alpha{p_2}_\beta{l_1}_\gamma(p_1^\mu-p_2^\mu) \nonumber\\
    &=&-4i(M\varepsilon^{\rho\sigma\mu\nu}({p_1}_{\rho}+{p_2}_{\rho}) {l_1}_\sigma + H_1\varepsilon^{\rho\sigma\mu\nu}{p_1}_{\rho}{p_2}_{\sigma}) \nonumber\\
    &&+\frac{2i}{M(\tau-1)}\left( |G_M|^2 - |G_E||G_M|\cos (\Delta \Phi)\right) \nonumber \\
    &&\times {p_1}_\alpha{p_2}_\beta{l_1}_\gamma(\varepsilon^{\alpha\beta\gamma\mu} Q^\nu-\varepsilon^{\alpha\beta\gamma\nu} Q^\mu) \nonumber \\
    &&+\frac{2}{M(\tau-1)}|G_E||G_M|\sin (\Delta \Phi) \nonumber \\
    &&\times {p_1}_\alpha{p_2}_\beta{l_1}_\gamma(\varepsilon^{\alpha\beta\gamma\mu} Q^\nu+\varepsilon^{\alpha\beta\gamma\nu} Q^\mu).
\end{eqnarray}
Among the terms in $H^{\mu\nu}_{S}$, only the last component is symmetric. Contracting $H^{\mu\nu}_{S}$ with $L_{\mu \nu}$, we obtain
\begin{eqnarray}
    H^{\mu\nu}_{S}L_{\mu \nu} &=& \frac{8Ep}{M(\tau-1)}|G_E||G_M|\cos\theta\sin (\Delta \Phi) \nonumber \\
    &&\times \varepsilon^{\alpha\beta\gamma\mu}{p_1}_\alpha{p_2}_\beta{l_1}_\gamma({k_1}_\mu-{k_2}_\mu).
\end{eqnarray}
To evaluate the term $\varepsilon^{\alpha\beta\gamma\mu}{p_1}_\alpha{p_2}_\beta{l_1}_\gamma({k_1}_\mu-{k_2}_\mu)$, it is convenient to switch to the rest frame of the $\Sigma^+$, where the four-momenta take the following forms: 
\begin{eqnarray}
    p_1^\prime &=& (M,0,0,0),  \\
    p_2^\prime&=&\left(\frac{E^2+p^2}{M},0,0,-\frac{2Ep}{M}\right), \\
    k_1^\prime &=& E\left(\frac{E-p\cos\theta}{M}, \sin \theta,0,\frac{E\cos\theta - p}{M} \right),  \\
    k_2^\prime&=&E\left(\frac{E+p\cos\theta}{M}, -\sin \theta,0,\frac{-E\cos\theta - p}{M} \right), \qquad
\end{eqnarray}
where $p_1^\prime$, $p_2^\prime$, $k_1^\prime$, and $k_2^\prime$ represent the four-momenta of the proton, antiproton, electron, and positron in the $\Sigma^+$ rest frame, respectively. By applying Lorentz invariance, we find that
\begin{eqnarray}
    &&\varepsilon^{\alpha\beta\gamma\mu}{p_1}_\alpha{p_2}_\beta{l_1}_\gamma({k_1}_\mu-{k_2}_\mu)  \nonumber \\
    &=& -M(\vec{p}_2^\prime\times(\vec{k}_1^\prime - \vec{k}_2^\prime))\cdot\vec{l}_1^\prime \nonumber \\
    &=& 4E^2p \sin \theta (\frac{1}{\sin \theta}\hat{p}\times\hat{k})\cdot \vec{l}_1^\prime.
\end{eqnarray}
After simplifying the expressions, the final result is
\begin{eqnarray}
    H^{\mu\nu}_{S}L_{\mu\nu} &=& sD(s)\sqrt{1-\alpha^2} \sin \theta \cos \theta \sin \Delta\Phi \nonumber \\
    &&\times \left[ \frac{1}{\sin \theta}(\hat{p}\times\hat{k})\cdot \vec{l}_1^\prime\right].
\end{eqnarray}

For the third component $H^{\mu\nu}_{\bar{S}}$, its explicit form is given by
\begin{eqnarray}
    H^{\mu\nu}_{\bar{S}}&=& \mathrm{Tr}\left[(\slashed{p}_1+M) \Gamma_\mu (\slashed{p}_2-M)\gamma^5 \left( \slashed{l}_2-H_2 \right)\bar{\Gamma}_\nu \right] \nonumber\\
    &=&-4i|G_M|^2(M\varepsilon^{\rho\sigma\mu\nu}({p_1}_{\rho}+{p_2}_{\rho}) {l_2}_\sigma - H_2\varepsilon^{\rho\sigma\mu\nu}{p_1}_{\rho}{p_2}_{\sigma}) \nonumber \\
    && -G_M\frac{F_2^*}{2M}4i\varepsilon^{\alpha\beta\gamma\mu}{p_1}_\alpha{p_2}_\beta{l_2}_\gamma(p_1^\nu-p_2^\nu) \nonumber\\
    &&+G_M^*\frac{F_2}{2M}4i\varepsilon^{\alpha\beta\gamma\nu}{p_1}_\alpha{p_2}_\beta{l_2}_\gamma(p_1^\mu-p_2^\mu) \nonumber\\
    &=&-4i(M\varepsilon^{\rho\sigma\mu\nu}({p_1}_{\rho}+{p_2}_{\rho}) {l_2}_\sigma - H_2\varepsilon^{\rho\sigma\mu\nu}{p_1}_{\rho}{p_2}_{\sigma}) \nonumber\\
    &&+\frac{2i}{M(\tau-1)}\left(|G_M|^2 - |G_E||G_M|\cos (\Delta \Phi)\right) \nonumber\\
    &&\times {p_1}_\alpha{p_2}_\beta{l_2}_\gamma(\varepsilon^{\alpha\beta\gamma\mu}Q^\nu-\varepsilon^{\alpha\beta\gamma\nu}Q^\mu) \nonumber\\
    && +\frac{2}{M(\tau-1)}|G_E||G_M|\sin (\Delta \Phi) \nonumber\\
    &&\times {p_1}_\alpha{p_2}_\beta{l_2}_\gamma(\varepsilon^{\alpha\beta\gamma\mu}Q^\nu+\varepsilon^{\alpha\beta\gamma\nu}Q^\mu).
\end{eqnarray}
By applying a similar approach, we derive the result for $H^{\mu\nu}_{\bar{S}}L_{\mu \nu}$:
\begin{eqnarray}
    H^{\mu\nu}_{\bar{S}}L_{\mu\nu} &=& sD(s)\sqrt{1-\alpha^2} \sin \theta \cos \theta \sin (\Delta\Phi) \nonumber\\
    &&\times \left[ \frac{1}{\sin \theta}(\hat{p}\times\hat{k})\cdot \vec{l}_2^\prime\right].
\end{eqnarray}

The final component, $H^{\mu\nu}_{S\bar{S}}$, is the most intricate to compute among the components discussed. It is expressed as:
\begin{eqnarray}
    H^{\mu\nu}_{S\bar{S}} &=& \mathrm{Tr}\left[(\slashed{p}_1+M) \gamma^5 \left( \slashed{l}_1+H_1 \right)\Gamma^\mu\right. \nonumber\\
    &&\left. \times (\slashed{p}_2-M) \gamma^5 \left( \slashed{l}_2-H_2 \right)\bar{\Gamma}^\nu \right] \nonumber\\
     &=& |G_M|^2A^{\mu\nu} +\frac{|F_2|^2}{4M^2}BQ^\mu Q^\nu  \nonumber\\
    &&-\frac{G_MF_2^*}{2M}T^{\mu\nu}_1 -\frac{G_M^*F_2}{2M}T^{\mu\nu}_2,
\end{eqnarray}
with
\begin{eqnarray}
    A^{\mu\nu}&=&\mathrm{Tr}\left[(\slashed{p}_1+M) \gamma^5 \left( \slashed{l}_1+H_1 \right)\gamma^\mu\right. \nonumber\\
    &&\left.\times (\slashed{p}_2-M) \gamma^5 \left( \slashed{l}_2-H_2 \right)\gamma^\nu \right] \nonumber \\
    &=&-4\left[ (l_1\cdot l_2+H_1H_2)(p_1^\mu p_2^\nu + p_1^\nu p_2^\mu - \frac{1}{2}sg^{\mu\nu})\right. \nonumber \\
    &&- P\cdot l_1(p_1^\mu l_2^\nu + p_1^\nu l_2^\mu) - P\cdot l_2 (p_2^\mu l_1^\nu + p_2^\nu l_1^\mu)   \nonumber \\
    && + \left.\frac{1}{2}s(l_1^\mu l_2^\nu + l_1^\nu l_2^\mu)  + P\cdot l_1 P\cdot l_2 g^{\mu\nu}  \right], \label{eq:A}
\end{eqnarray}
\begin{eqnarray}
    B &=& \mathrm{Tr}\left[(\slashed{p}_1+M) \gamma^5 \left( \slashed{l}_1+H_1 \right) (\slashed{p}_2-M) \gamma^5 \left( \slashed{l}_2-H_2 \right)\right] \nonumber\\
    &=&  4\left[ (-\frac{1}{2}s + 2M^2)l_1 \cdot l_2 + (\frac{1}{2}s + 2M^2)H_1H_2 \right. \nonumber\\
    &&+\left.P\cdot l_1 P\cdot l_2  - 2M(H_1P\cdot l_2 + H_2 P \cdot l_1)\right] \label{eq:B},
\end{eqnarray}
and
\begin{eqnarray}
    T_1^{\mu\nu} &=& \mathrm{Tr}\left[(\slashed{p}_1+M) \gamma^5 \left( \slashed{l}_1+H_1 \right)\gamma^\mu\right. \nonumber\\
    &&\times\left.(\slashed{p}_2-M) \gamma^5 \left( \slashed{l}_2-H_2 \right) \right]Q^\nu \nonumber\\
    &=&4\left[(Ml_1\cdot l_2 - H_2p_2\cdot l_1)p_1^\mu  \right.\nonumber\\
    &&- \left.(Ml_1\cdot l_2 - H_1p_1\cdot l_2)p_2^\mu \right. \nonumber \\
    &&- \left.(Mp_1\cdot l_2 -H_2p_1\cdot p_2)l_1^\mu \right. \nonumber \\
    &&+ \left.(Mp_2\cdot l_1 -H_1p_1\cdot p_2)l_2^\mu\right]Q^\nu,
\end{eqnarray}
\begin{eqnarray}
    T_2^{\mu\nu} &=& \mathrm{Tr}\left[(\slashed{p}_1+M) \gamma^5 \left( \slashed{l}_1+H_1 \right)\right. \nonumber \\
    &&\left.(\slashed{p}_2-M) \gamma^5 \left( \slashed{l}_2-H_2 \right)\gamma^\nu \right]Q^\mu  \nonumber\\
    &=&4\left[ (Ml_1\cdot l_2 - H_2p_2\cdot l_1)p_1^\nu  \right. \nonumber\\
    &&- \left.(Ml_1\cdot l_2 - H_1p_1\cdot l_2)p_2^\nu  \right. \nonumber\\
    &&- \left.(Mp_1\cdot l_2 -H_2 p_1\cdot p_2)l_1^\nu \right. \nonumber\\
    &&+ \left.(Mp_2\cdot l_1 -H_1 p_1\cdot p_2)l_2^\nu \right]Q^\mu,
\end{eqnarray}
Substituting these into the expression for $H^{\mu\nu}_{S\bar{S}}$, the terms involving $T_1^{\mu\nu}$ and $T_2^{\mu\nu}$ can be expressed as:
\begin{eqnarray}
    &&-\frac{G_MF_2^*}{2M}T^{\mu\nu}_1 -\frac{G_M^*F_2}{2M}T^{\mu\nu}_2 \nonumber \\
    &=&\frac{2}{M(\tau -1)}\left(|G_M|^2-|G_E||G_M|\cos (\Delta \Phi)\right)T_S^{\mu\nu} \nonumber\\
    &&+\frac{2i|G_E||G_M|}{M(\tau -1)}\sin (\Delta \Phi)T_A^{\mu\nu}.
\end{eqnarray}
The corresponding $T_S^{\mu\nu}$ and $T_A^{\mu\nu}$ are defined as:
\begin{eqnarray}
    T_S^{\mu\nu} &=& \frac{1}{4}(T_1^{\mu\nu} + T_2^{\mu\nu}) \nonumber \\
    &=& (H_2P\cdot l_1 + H_1P\cdot l_2)(p_1^\mu p_2^\nu + p_1^\nu p_2^\mu) \nonumber \\
    && - 2H_2P\cdot l_1 p_1^\mu p_1^\nu - 2H_1P\cdot l_2 p_2^\mu p_2^\nu \nonumber \\
    && + (MP\cdot l_1 - \frac{1}{2}sH_1)(Q^\mu l_2^\nu + Q^\nu l_2^\mu ) \nonumber \\
    && - (MP\cdot l_2 - \frac{1}{2}sH_2)(Q^\mu l_1^\nu + Q^\nu l_1^\mu ) \nonumber \\
    && + 2M(l_1\cdot l_2+H_1H_2)Q^\mu Q^\nu,  \label{eq:T}
\end{eqnarray}
\begin{eqnarray}
    T_A^{\mu\nu} &=& \frac{1}{4}(T_1^{\mu\nu} - T_2^{\mu\nu}) \nonumber \\
    &=& (H_2P\cdot l_1 - H_1P\cdot l_2) (p_1^\mu p_2^\nu - p_1^\nu p_2^\mu)  \nonumber\\
    &&+ (MP\cdot l_2 - \frac{1}{2}sH_2)(Q^\mu l_1^\nu - Q^\nu l_1^\mu ) \nonumber \\
    &&- (MP\cdot l_1 - \frac{1}{2}sH_1)(Q^\mu l_2^\nu - Q^\nu l_2^\mu ). 
\end{eqnarray}
It can be observed that $T_A^{\mu\nu}$ is an antisymmetric tensor, which implies that $T_A^{\mu\nu}L_{\mu\nu} = 0$.

The next step involves the contraction of $A^{\mu\nu}$ and $T_S^{\mu\nu}$ with $L_{\mu\nu}$. However, before proceeding with this operation, it is necessary to compute some fundamental components to simplify the calculation and facilitate the derivation of the final result. As previously mentioned, $l_1'$ and $l_2'$ represent the four-momenta in the $\Sigma^+$ and $\bar{\Sigma}^-$ center-of-mass frames, respectively. To perform the transformation into the $e^+e^-$ center-of-mass frame, we obtain:
\begin{eqnarray}
    l_1 &=& \frac{H_1}{M}(E,\vec{p}) + \frac{\eta_1}{M}(p, E\hat{p}) + (0, \vec{l}_{1\perp}^\prime),\\
    l_2 &=& \frac{H_2}{M}(E,-\vec{p}) + \frac{\eta_2}{M}(-p, E\hat{p}) + (0, \vec{l}_{2\perp}^\prime),
\end{eqnarray}
where
\begin{eqnarray}
    \eta_1 = \hat{p}\cdot \vec{l}_1^\prime, \qquad  \eta_2 = \hat{p}\cdot \vec{l}_2^\prime,
\end{eqnarray}
and
\begin{eqnarray}
    \vec{l}_{1\perp}^\prime &=& \vec{l}_1^\prime - \eta_1\hat{p}, \\
    \vec{l}_{2\perp}^\prime &=& \vec{l}_2^\prime - \eta_2\hat{p}.
\end{eqnarray}
In addition, we also have defined the following shorthand
\begin{eqnarray}
    \zeta_1 = \vec{l}_{1\perp}^\prime \cdot \hat{k},\qquad \zeta_2 = \vec{l}_{2\perp}^\prime \cdot \hat{k},\qquad  \gamma =\vec{l}_{1\perp}^\prime \cdot \vec{l}_{2\perp}^\prime.
\end{eqnarray}
With these definitions in place, we can proceed to derive the following results.
\begin{eqnarray}
    l_1\cdot l_2 &=& \frac{H_1H_2}{M^2}(E^2+p^2)-\frac{2E^2}{M^2}\eta_1\eta_2 +\frac{2EH_2}{M^2}p\eta_1  \nonumber \\
    &&- \frac{2EH_1}{M^2}p\eta_2 - (\gamma - \eta_1\eta_2), \label{eq:ll}
\end{eqnarray}
\begin{eqnarray}
    P\cdot l_1 &=&\frac{2E^2H_1}{M} +  \frac{2E}{M}p\eta_1, \\
    P\cdot l_2 &=&\frac{2E^2H_2}{M} - \frac{2E}{M}p\eta_2,
\end{eqnarray}
and
\begin{eqnarray}
    l_1\cdot k_1 &=&\frac{EH_1}{M}(E-p\cos \theta) - E\zeta_1 \nonumber\\
    &&+ \frac{E}{M}(p-E\cos \theta)\eta_1, \\
    l_1 \cdot k_2 &=& \frac{EH_1}{M}(E+p\cos \theta)   + E\zeta_1 \nonumber\\
    &&+ \frac{E}{M}(p+E\cos \theta)\eta_1, \\
    l_2\cdot k_1 &=&\frac{EH_2}{M}(E+p\cos \theta)  - E\zeta_2  \nonumber\\
    &&- \frac{E}{M}(p+E\cos \theta)\eta_2,  \\
    l_2 \cdot k_2 &=& \frac{EH_2}{M}(E-p\cos \theta)  + E\zeta_2 \nonumber\\
    &&- \frac{E}{M}(p-E\cos \theta)\eta_2.
\end{eqnarray}
Using these results, we can readily obtain
\begin{eqnarray}
    &&(l_1^\mu l_2^\nu + l_1^\nu l_2^\mu)L_{\mu \nu} \nonumber\\
    &=&s(\gamma-\zeta_1\zeta_2)- \frac{sE}{M}(\eta_2\zeta_1+\eta_1\zeta_2)\cos \theta \nonumber\\
    &&-\frac{sH_1H_2}{M^2}p^2\sin^2\theta + \frac{sE^2}{M^2}\eta_1\eta_2\sin^2 \theta  \nonumber \\
    && - \frac{sEH_2p}{M^2}\eta_1\sin^2\theta  + \frac{sEH_1p}{M^2}\eta_2\sin^2\theta \nonumber \\
    && +\frac{sH_2}{M}p\zeta_1 \cos \theta - \frac{sH_1}{M}p\zeta_2 \cos \theta, 
\end{eqnarray}
and
\begin{eqnarray}
    (p_1^\mu l_1^\nu + p_1^\nu l_1^\mu)L_{\mu \nu} &=&   \frac{sH_1}{M}p^2\sin^2\theta + \frac{sE}{M}p\eta_1\sin^2\theta \nonumber\\
    &&- sp\zeta_1\cos \theta, \\
    (p_2^\mu l_2^\nu + p_2^\nu l_2^\mu)L_{\mu \nu} &=&   \frac{sH_2}{M}p^2\sin^2\theta - \frac{sE}{M}p\eta_2\sin^2\theta \nonumber\\
    &&+ sp\zeta_2\cos \theta.
\end{eqnarray}
Furthermore, due to the condition $(p_1^{\mu}+p_2^\mu)L_{\mu\nu} = 0$, one arrive at
\begin{eqnarray}
    (p_1^\mu l_1^\nu + p_1^\nu l_1^\mu)L_{\mu \nu} &=& - (p_2^\mu l_1^\nu + p_2^\nu l_1^\mu)L_{\mu \nu},\\
    (p_2^\mu l_2^\nu + p_2^\nu l_2^\mu)L_{\mu \nu} &=& - (p_1^\mu l_2^\nu + p_1^\nu l_2^\mu)L_{\mu \nu}. \label{eq:pl}
\end{eqnarray}
Substituting Eqs.~\eqref{eq:ll} and~\eqref{eq:pl} into $A^{\mu\nu}$, $B$, and $T_S^{\mu\nu}$, we find
\begin{eqnarray}
    A^{\mu\nu}L_{\mu\nu} &=& 4\left[2sE^2(\eta_1\eta_2 + \zeta_1\zeta_2) -2sE^2\eta_1\eta_2\sin^2\theta  \right. \nonumber\\
    && + 2sEM(\eta_2\zeta_1+\eta_1\zeta_2)\cos \theta \nonumber\\
    &&- \left.sp^2(\gamma - \eta_1\eta_2)\sin^2\theta \right] , \label{eq:AL}
\end{eqnarray}
\begin{eqnarray}
    B=8p^2(\gamma -  \eta_1\eta_2), \label{eq:BB}
\end{eqnarray}
and
\begin{eqnarray}
    T_S^{\mu\nu}L_{\mu\nu} &=& -4sp^2\left[M(\gamma - \eta_1\eta_2)\sin^2\theta \right. \nonumber\\
    &&+\left. E(\eta_2\zeta_1+\eta_1\zeta_2)\cos\theta \right]. \label{eq:TL}
\end{eqnarray}
Finally, substituting Eqs.~\eqref{eq:AL} and~\eqref{eq:TL} into $H^{\mu\nu}_{S\bar{S}}$ yields
\begin{eqnarray}
    H_{55}^{\mu\nu}L_{\mu\nu} &=& sD(s)\left[(\alpha + \cos^2\theta)\eta_1\eta_2 + (1+\alpha)\zeta_1\zeta_2 \right. \nonumber\\
    && + \sqrt{1-\alpha^2}\cos \theta \cos(\Delta \Phi)(\eta_1\zeta_2+\eta_2\zeta_1) \nonumber\\
    &&- \left.\alpha \gamma \sin^2\theta \right].
\end{eqnarray}

Since the total decay widths of $\Sigma^+$ and $\bar{\Sigma}^-$ are relatively small, we can apply the narrow-width approximation to $|D_1|^2$ and $|D_2|^2$:
\begin{eqnarray}
    |D_1|^2 = \frac{\pi}{M\Gamma}\delta(s_1-M^2),\\
    |D_2|^2 = \frac{\pi}{M\Gamma}\delta(s_2-M^2).
\end{eqnarray}
Therefore, the cross section of the process  $e^+e^-\to \Sigma^+(\to p \pi^0)\bar{\Sigma}^-(\to \bar{p} \pi^0)$ can be reduced to
\begin{eqnarray}
    \mathrm{d}\sigma &=& \frac{1}{2s}\frac{(4\pi\alpha_{em})^2}{s^2}\frac{\pi^2}{M^2\Gamma^2}\delta(s_1-M^2)\delta(s_2-M^2) \nonumber \\
    &&\times H^{\mu\nu}L_{\mu\nu}\mathrm{d}\Phi_4 \nonumber\\
    &=&\frac{\pi\beta\alpha_{em}^2}{s^2}D(s)\left(1+\frac{\alpha}{3}\right)\frac{1}{\Gamma^2}\frac{R|\vec{l}_1^\prime|}{8\pi M}\frac{\bar{R}|\vec{l}_2^\prime|}{8\pi M} \nonumber\\
    &&\times W(\Omega,\Omega_p,\Omega_{\bar{p}})\,\mathrm{d} \Omega\mathrm{d} \Omega_p \mathrm{d} \Omega_{\bar{p}}  \nonumber\\
    &=&\sigma(e^+e^-\to \Sigma^+\bar{\Sigma}^-)\mathrm{Br}(\Sigma^+\to p \pi^0)\mathrm{Br}(\bar{\Sigma}^-\to \bar{p} \pi^0) \nonumber\\
    &&\times W(\Omega,\Omega_p,\Omega_{\bar{p}})\,\mathrm{d} \Omega\mathrm{d} \Omega_p \mathrm{d} \Omega_{\bar{p}}, 
\end{eqnarray}
where
\begin{eqnarray}
    \sigma(e^+e^-\to \Sigma^+\bar{\Sigma}^-) &=& \frac{\pi\beta\alpha_{em}^2}{s^2}D(s)\left(1+\frac{\alpha}{3}\right)
\end{eqnarray}
is the total cross section of the reaction $e^+e^- \to \Sigma^+ \bar{\Sigma}^-$, as in Eq.~\ref{eq:cs1}, except that the Sommerfeld–Gamow factor $C$ is omitted here. Here, $\beta$ represents the velocity of the $\Sigma^+$ hyperon in the center-of-mass frame, and $\mathrm{Br}(\Sigma^+ \to p \pi^0)$ and $\mathrm{Br}(\bar{\Sigma}^- \to \bar{p} \pi^0)$ denote the branching ratios for $\Sigma^+ \to p \pi^0$ and $\bar{\Sigma}^- \to \bar{p} \pi^0$, respectively.
\begin{eqnarray}
    \mathrm{Br}(\Sigma^+\to p \pi^0) &=& \frac{1}{\Gamma}\frac{R|\vec{l}_1^\prime|}{8\pi M},\\
    \mathrm{Br}(\bar{\Sigma}^-\to \bar{p} \pi^0)&=& \frac{1}{\Gamma}\frac{\bar{R}|\vec{l}_2^\prime|}{8\pi M}.
\end{eqnarray}

In addition, the angular distribution function $W(\Omega, \Omega_p, \Omega_{\bar{p}})$ represents the angular distribution characteristics of this reaction. For brevity, we introduce the variable $\xi = (\Omega, \Omega_p, \Omega_{\bar{p}})$. The explicit form of the function is given by
\begin{eqnarray}
    W(\xi)&=&\frac{1}{N_{\xi}}\left[
    1 + \alpha \mathcal{T}_5(\xi)   + \alpha_{\Sigma^+}\bar{\alpha}_{\bar{\Sigma}^-}\left(\mathcal{T}_1(\xi) + \alpha \mathcal{T}_6(\xi)\right)\right.
    \nonumber \\
    &&+ \sqrt{1-\alpha^2}\sin (\Delta\Phi)\left(\alpha_{\Sigma^+} \mathcal{T}_3(\xi) +\bar{\alpha}_{\bar{\Sigma}^-} \mathcal{T}_4(\xi) \right)  \nonumber\\
    &&+ \left.\alpha_{\Sigma^+}\bar{\alpha}_{\bar{\Sigma}^-}\sqrt{1-\alpha^2}\cos(\Delta\Phi)\mathcal{T}_2(\xi)  \right], \label{eq:wxi}
\end{eqnarray}
where the $N_\xi$ is a normalization constant
\begin{eqnarray}
    N_\xi = 64\pi^3\left(1+\frac{\alpha}{3}\right),
\end{eqnarray}
such that
\begin{eqnarray}
    \int W(\Omega,\Omega_p,\Omega_{\bar{p}})\,\mathrm{d} \Omega\mathrm{d} \Omega_p \mathrm{d} \Omega_{\bar{p}} = 1.
\end{eqnarray}

The angular functions $\mathcal{T}_i(\xi)$ $i=(1, 2, \cdots, 6)$ are given as follows:
\begin{eqnarray}
    \mathcal{T}_1(\xi)&=&\sin^2 \theta \sin \theta_p \cos \phi_p\sin \theta_{\bar{p}}  \cos \theta_{\bar{p}} + \cos^2 \theta \cos \theta_p \cos \theta_{\bar{p}}, \nonumber \\
    \mathcal{T}_2(\xi)&=&\sin \theta \cos \theta ( \cos \theta_p \sin \theta_{\bar{p}} \cos \phi_{\bar{p}}  + \cos \theta_{\bar{p}} \sin \theta_p  \cos \phi_p ),  \nonumber \\
    \mathcal{T}_3(\xi)&=&\sin \theta \cos \theta \sin \theta_p \sin \phi_p, \nonumber \\
    \mathcal{T}_4(\xi)&=&\sin \theta \cos \theta \sin \theta_{\bar{p}} \sin \phi_{\bar{p}}, \nonumber \\
    \mathcal{T}_5(\xi)&=&\cos^2 \theta,  \nonumber \\
    \mathcal{T}_6(\xi)&=&\cos \theta_p \cos \theta_{\bar{p}} - \sin^2 \theta \sin \theta_p \sin \phi_p \sin \theta_{\bar{p}} \sin \phi_{\bar{p}}. \nonumber   
\end{eqnarray}
It is worth mentioning that if the $y$ axis direction is defined by $\vec{k}\times \vec{p}$, The $\sin \theta$ in the angular distribution function $\mathcal{T}_i(\xi)$ needs to be replaced with $-\sin \theta$.

With the normalized angular distribution function $W(\xi)$, we can calculate the mean value of the proton and antiproton momentum direction. The only non-zero terms are
\begin{eqnarray}
    W_{1y} &=& \int \sin \theta_p \sin \phi_p W\, \mathrm{d} \phi \mathrm{d}\Omega_p \mathrm{d} \Omega_{\bar{p}} \nonumber \\
    &=&\frac{\alpha_{\Sigma^+}}{2(3+\alpha)}\sqrt{1-\alpha^2}\sin (\Delta\Phi)\sin \theta \cos \theta,\\
    W_{2y} &=& \int \sin \theta_{\bar{p}} \sin \theta_{\bar{p}} W \, \mathrm{d} \phi \mathrm{d}\Omega_p \mathrm{d} \Omega_{\bar{p}}  \nonumber \\
    &=&\frac{\alpha_{\bar{\Sigma}^-}}{2(3+\alpha)}\sqrt{1-\alpha^2}\sin (\Delta\Phi)\sin \theta \cos \theta, \\
    W_{zx} &=& \int \cos \theta_p \sin \theta_{\bar{p}}  \cos \theta_{\bar{p}} W\, \mathrm{d} \phi \mathrm{d}\Omega_p \mathrm{d} \Omega_{\bar{p}}  \nonumber\\
    &=&\frac{\alpha_{\Sigma^+}\bar{\alpha}_{\bar{\Sigma}^-}}{6(3+\alpha)}\sqrt{1-\alpha^2}\cos (\Delta\Phi)\sin \theta \cos \theta,
\end{eqnarray}
and
\begin{eqnarray}
    W_{xx} &=&  \int \sin \theta_p \cos \phi_p\sin \theta_{\bar{p}}  \cos \theta_{\bar{p}} W\, \mathrm{d} \phi \mathrm{d}\Omega_p \mathrm{d} \Omega_{\bar{p}}  \nonumber \\
    &=&\frac{\alpha_{\Sigma^+}\bar{\alpha}_{\bar{\Sigma}^-}}{6(3+\alpha)}\sin^2\theta ,\\
    W_{yy}&=& \int \sin \theta_p \sin \phi_p\sin \theta_{\bar{p}}  \sin \phi_{\bar{p}} W\, \mathrm{d} \phi \mathrm{d}\Omega_p \mathrm{d} \Omega_{\bar{p}}  \nonumber \\
    &=&-\frac{\alpha_{\Sigma^+}\bar{\alpha}_{\bar{\Sigma}^-}}{6(3+\alpha)}\alpha \sin^2\theta, \\
    W_{zz}&=& \int \cos \theta_p \cos \theta_{\bar{p}}   W\, \mathrm{d} \phi \mathrm{d}\Omega_p \mathrm{d} \Omega_{\bar{p}}  \nonumber \\
    &=&\frac{\alpha_{\Sigma^+}\bar{\alpha}_{\bar{\Sigma}^-}}{6(3+\alpha)}\left(\alpha + \cos^2\theta \right).
\end{eqnarray}
From the above equations, we can find that the distributions of $W_{xx}$, $W_{yy}$, $W_{zz}$ depend only on the modulus of the EMFFs, but $W_{1y}$, $W_{2y}$ and $W_{zx}$ are related to the relative phase of the EMFFs in addition to the $\alpha$. This allows us to extract the relative phase $\Delta \Phi$ of the EMFFs by measuring $W_{1y}$ and $W_{zx}$. The following observable quantity is commonly measured in experiments, which is defined as
\begin{eqnarray}
    M(\cos \theta_k) = \frac{m_{\theta}}{N}\sum^{N_k}_i[\sin \theta_p^i \sin \phi_p^i - \sin \theta_{\bar{p}}^i \sin \phi_{\bar{p}}^i], \label{eq:mk}
\end{eqnarray}
Here, $m_\theta$ denotes the number of bins for $\cos \theta$, $N$ represents the total number of events, and $N_k$ refers to the number of events in the $k$th bin. To extend this formulation to the continuous limit where $m_\theta \to +\infty$, we obtain
\begin{eqnarray}
    M(\cos \theta)&=&\frac{m_\theta}{N}\times N\frac{2}{m_\theta} \int \mathrm{d} \phi \mathrm{d}\Omega_p \mathrm{d} \Omega_{\bar{p}}\  \left[W \right. \nonumber\\
    &&\times \left.(\sin \theta_p \sin \phi_p-\sin \theta_{\bar{p}} \sin \phi_{\bar{p}})\right]  \nonumber \\
    &=&\frac{\alpha_{\Sigma^+}-\alpha_{\bar{\Sigma}^-}}{3+\alpha}\sqrt{1-\alpha^2}\sin(\Delta \Phi)\sin\theta \cos \theta \nonumber\\
    &=&2\left(W_{1y}-W_{2y}\right).
\end{eqnarray}
We would like to emphasize again that the results of this physical quantity are dependent on the choice of the reference frame, and in particular that different selections can differ by plus or minus signs.

\begin{figure}[htbp]
    \centering
    \includegraphics[width=0.9\linewidth]{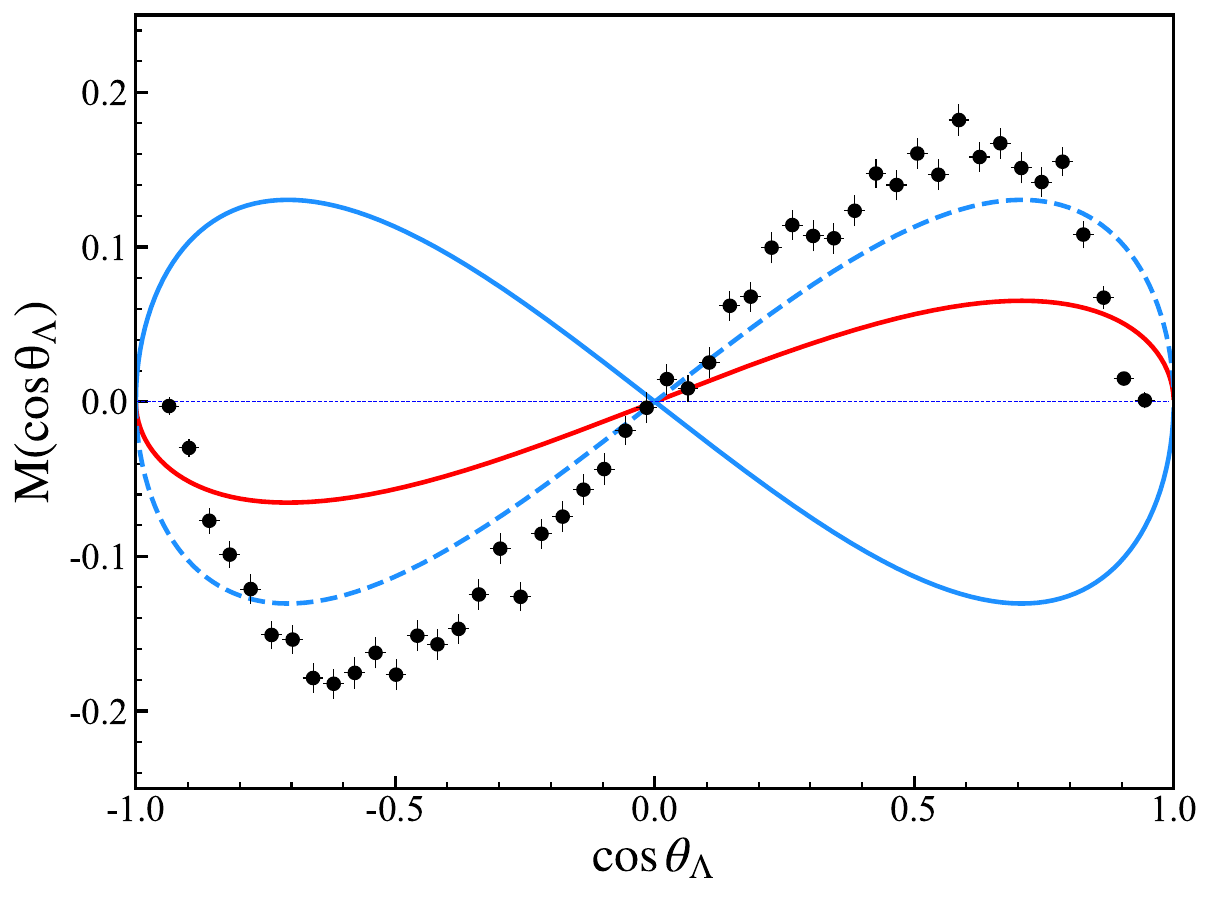}

    \caption{The moment $M(\cos_\Lambda)$ of $\Lambda$ hyperon. The red solid line represents the formula Eq.~\eqref{eq:mlambda}, while the blue solid line is based on Eq.~\eqref{eq:mlam2}, and the blue dashed line is the opposite of the blue solid line. The data are taken from BESIII~\cite{BESIII:2018cnd}.}
    \label{fig:MLam}
\end{figure}

In this context, we draw attention to several issues in some current experimental articles. Specifically, in Refs.~\cite{BESIII:2018cnd} and ~\cite{BESIII:2022qax}, the authors presented the definition of moments as given in Eq.~\ref{eq:mk}, but when extended to the continuous case, they present 
\begin{eqnarray}
    \tilde{M}(\cos \theta_\Lambda) &=& \frac{\alpha_--\alpha_+}{2}\frac{1}{(3+\alpha)}\sqrt{1-\alpha^2} \nonumber \\
    &&\times \sin(\Delta \Phi)\sin\theta_\Lambda \cos \theta_\Lambda, \label{eq:mlambda}
\end{eqnarray}
where $\theta_\Lambda$ represents the scattering angle between the $\Lambda$ momentum and the electron momentum. However, their definition of $\hat{y}$ as the direction of $\vec{k}\times \vec{p}$ leads to an inconsistency. Therefore, Eq.~\ref{eq:mlambda} is incorrect and should be revised to
\begin{eqnarray}
    M(\cos \theta_\Lambda) &=& -\frac{\alpha_--\alpha_+ }{3+\alpha} \sqrt{1-\alpha^2}\nonumber \\
    &&\times \sin(\Delta \Phi)\sin\theta_\Lambda \cos \theta_\Lambda. \label{eq:mlam2}
\end{eqnarray}
The results obtained by using the parameters from Ref.~\cite{BESIII:2018cnd} in Eq.~\ref{eq:mlam2} disagree with the experimental data, as illustrated in Fig.~\ref{fig:MLam}. In this figure, the black dot represents the experimental data from Ref.~\cite{BESIII:2018cnd}, while the red solid line corresponds to Eq.~\ref{eq:mlambda} and the blue line represents Eq.~\ref{eq:mlam2}. The blue dashed line is the negative of the blue solid line. As shown, the red line does not match the experimental data in magnitude. Although the blue line's magnitude aligns well with the experimental results, its behavior is completely opposite. This discrepancy may arise from the coordinate system definition in the literature or could indicate that the extracted $\Delta \Phi$ should actually be $-\Delta \Phi$.

\begin{figure}[htbp]
    \centering
    \includegraphics[width=0.9\linewidth]{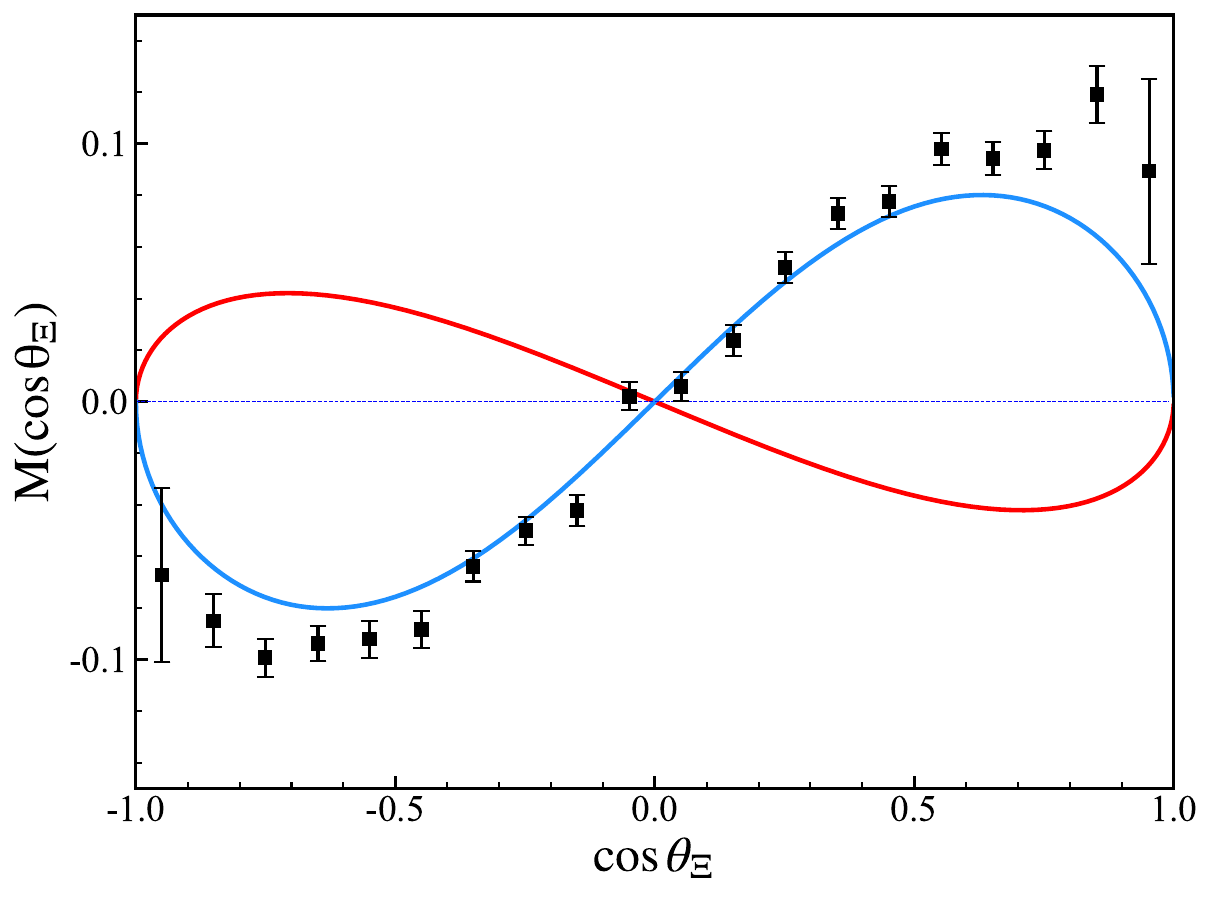}

    \caption{The moment $M(\cos_\Xi)$ of $\Xi^0$ hyperon. The red solid line represents the formula Eq.~\eqref{eq:mxi}, while the blue solid line is based on Eq.~\eqref{eq:mxi2}. The data are taken from BESIII~\cite{BESIII:2023drj}.}
    \label{fig:MXi}
\end{figure}

Furthermore, we examine another study~\cite{BESIII:2023drj}, where the authors investigated the reaction $e^+e^-\to \Xi^0(\to \Lambda \pi^0)\bar{\Xi}^0(\to \bar{\Lambda}\pi^0)$. In this work, they provided the following definition for the moments:
\begin{eqnarray}
    && M^k(\cos \theta_\Xi) =  \nonumber \\
    && \frac{1}{N_k}\sum^{N_k}_i[\sin \theta_\Lambda^i \sin \phi_\Lambda^i - \sin \theta_{\bar{\Lambda}}^i \sin \phi_{\bar{\Lambda}}^i], \label{eq:mxik}
\end{eqnarray}
This definition differs from the one presented in Eq.~\eqref{eq:mk}. In their analysis, $\theta_\Xi$ represents the angle between the $\Xi^0$ momentum $\vec{p}$ and the positron momentum $-\vec{k}$, with the y-axis defined as $-\vec{k}\times \vec{p}$. Despite these differences in the definition of $M$ and the reference system, the authors derive the same expression for $M$,
\begin{eqnarray}
      \tilde{M}(\cos \theta_\Xi) &=& \frac{\alpha_\Xi- \bar{\alpha}_\Xi}{2}\frac{1}{3+\alpha }\sqrt{1-\alpha^2}  \nonumber \\ 
    &&\times \sin(\Delta \Phi)\sin \theta_\Xi \cos \theta_\Xi, \label{eq:mxi}
\end{eqnarray}
which is incorrect. Since the factor in front of the summation sign in Eq.~\ref{eq:mxik} is $1/N_k$, the moments $ M(\cos \theta_\Xi)$ should be modified to
\begin{eqnarray}
    M(\cos \theta_\Xi) &=& -\frac{\alpha_\Xi- \bar{\alpha}_\Xi}{3(1+\alpha \cos ^2 \theta)}\sqrt{1-\alpha^2}  \nonumber \\ 
    &&\times \sin(\Delta \Phi)\sin \theta_\Xi \cos \theta_\Xi. \label{eq:mxi2}
\end{eqnarray}
In Fig.~\ref{fig:MXi}, we compare the line shapes of these two expressions using the parameters from Ref.~\cite{BESIII:2023drj}. The solid red curve represents $\tilde{M}(\cos \theta_\Xi)$, while the dashed blue curve represents $M(\cos \theta_\Xi)$. It is evident that $\tilde{M}(\cos \theta_\Xi)$ does not agree with the experimental data, whereas $M(\cos \theta_\Xi)$ is consistent with the experimental results.

Next, we focus on the spin distribution of $\Sigma$ hyperons produced in the $e^+e^- \to \Sigma \bar{\Sigma}$ reaction without considering the decay process of $\Sigma$. For the final-state particles $\Sigma$ and $\bar{\Sigma}$ with spins $s_1$ and $s_2$, the amplitude is expressed as
\begin{eqnarray}
    \mathcal{M}_{s_1s_2} = \frac{e^2}{q^2}\bar{v}(k_2)\gamma^\mu u(k_1)u(p_1,s_1)\Gamma_\mu v(p_2,s_2),
\end{eqnarray}
where the definition of the momentum and coordinate system are defined as previously described, and $s_1$ and $s_2$ are the spin four-vector of $\Sigma$ and $\bar{\Sigma}$ hyperon. Assuming the spin directions of $\Sigma$ and $\bar{\Sigma}$ in their respective center-of-mass systems are $\vec{n}_1$ and $\vec{n}_2$, we define their spin four-vectors as

\begin{eqnarray}
    s_1(\vec{p}_1,\vec{n}_1) = \frac{\vec{n}_1\cdot\hat{p}_1}{M}(|\vec{p}_1|, E\hat{p}_1)+(0,\vec{n}_{1\perp}), \\
     s_2(\vec{p}_2,\vec{n}_2) = \frac{\vec{n}_2\cdot\hat{p}_2}{M}(|\vec{p}_2|, E\hat{p}_2)+(0,\vec{n}_{2\perp}).
\end{eqnarray}
Here,
\begin{eqnarray}
    \vec{n}_1&=&(\sin \theta_1 \cos\phi_1, \sin \theta_1 \sin \phi_1, \cos \theta_1), \\
    \vec{n}_2&=&(\sin \theta_2 \cos\theta_2, \sin \theta_2 \sin \theta_2, \cos \theta_2),
\end{eqnarray}
and  $\vec{n}_{1\perp} = \vec{n}_1-\hat{p}_1(\vec{n}_1\cdot\hat{p}_1)$, $\vec{n}_{2\perp} = \vec{n}_2-\hat{p}_2(\vec{n}_2\cdot\hat{p}_2)$ and define $(\theta_1,\phi_1)=\Omega_1$, $(\theta_2,\phi_2)=\Omega_2$.
Besides, one can find the four-momentum and four-vector are orthogonal, 
\begin{eqnarray}
    p_1\cdot s_1=0, \qquad p_2\cdot s_2 = 0.
\end{eqnarray}

Then, the differential cross section for the reaction is given by
\begin{eqnarray}
    \mathrm{d}\sigma_{s_1s_2} &=& \frac{1}{4}\frac{1}{2s}|\mathcal{M}_{s_1s_2}|^2\frac{p}{16\pi^2\sqrt{s}}\ \mathrm{d} \Omega  \nonumber \\
    &=&\frac{\alpha_{em}^2\beta}{4s^3}G^{\mu\nu}_{s_1s_2}L_{\mu\nu} \ \mathrm{d} \Omega,
\end{eqnarray}
where the tensor $G^{\mu\nu}_{s_1s_2}$ is defined as
\begin{eqnarray}
    G^{\mu\nu}_{s_1s_2} &=&\bar{u}(p_1,s_1)\Gamma^\mu v(p_2,s_2) \bar{v}(p_2,s_2) \bar{\Gamma}^\nu  u(p_1,s_1) \nonumber \\
    &=&\mathrm{Tr}\left[ (\slashed{p}_1+M)\frac{1}{2}(1+\gamma^5\slashed{s}_1)\Gamma^\mu \right. \nonumber\\
    &&\left.\times (\slashed{p}_2-M)\frac{1}{2}(1-\gamma^5\slashed{s}_2)\bar{\Gamma}^\nu\right].
\end{eqnarray}
The angular distribution function $W_{s_1s_2}$ is derived in a similar manner to the previous approach, which yields
\begin{eqnarray}
    W_{s_1s_2} &=& \frac{1}{N_s}G^{\mu\nu}_{s_1s_2}L_{\mu\nu} \nonumber\\
    &=& \frac{1}{16\pi \left(1+\frac{\alpha}{3}\right)}\left[
    1 + \alpha \mathcal{F}_5   + \mathcal{F}_1 + \alpha \mathcal{F}_6\right.
    \nonumber \\
    &&+ \sqrt{1-\alpha^2}\sin (\Delta\Phi)\left(\mathcal{F}_3 + \mathcal{F}_4 \right)  \nonumber\\
    &&+ \left.\sqrt{1-\alpha^2}\cos(\Delta\Phi)\mathcal{F}_2  \right] \label{eq:wxi},
\end{eqnarray}
where the auxiliary function $\mathcal{F}_i$ $i=(1, 2, \cdots, 6)$ is given by
\begin{eqnarray}
    \mathcal{F}_1&=&\sin^2 \theta \sin \theta_1 \sin \theta_2 \cos \phi_1 \cos \phi_2 + \cos^2 \theta \cos \theta_1 \cos \theta_2 ,\nonumber\\
    \mathcal{F}_2&=&\sin \theta \cos \theta (\cos \theta_2 \sin \theta_1  \cos \phi_1 + \cos \theta_1 \sin \theta_2 \cos \phi_2) ,\nonumber\\
    \mathcal{F}_3&=&\sin \theta \cos \theta \sin \theta_1 \sin \phi_1 ,\nonumber\\
    \mathcal{F}_4&=&\sin \theta \cos \theta \sin \theta_2 \sin \phi_2 ,\nonumber\\
    \mathcal{F}_5&=&\cos^2 \theta ,\nonumber\\
    \mathcal{F}_6&=&\cos \theta_1 \cos \theta_2 - \sin^2 \theta \sin \theta_1 \sin \theta_2 \sin \phi_1 \sin \phi_2  .\nonumber
\end{eqnarray}
$N_s$ is a normalization constant, such that
\begin{eqnarray}
     \frac{1}{N_s}\sum_{\pm s_1\pm s_2}\int G^{\mu\nu}_{s_1s_2}L_{\mu\nu}\, \mathrm{d}\Omega = 1.
\end{eqnarray}
It can be observed that $W(\xi)$ and $W_{s_1s_2}$ share the same algebraic structure, with only a few constant factors differing between them. Therefore, the momentum distribution of the proton in the $e^+e^- \to \Sigma^+(\to p \pi^0)\bar{\Sigma}^-(\to \bar{p} \pi^0)$ process can reflect the spin distribution of the $e^+e^- \to \Sigma^+ \bar{\Sigma}^-$ reaction.

Finally, the differential cross section is given by
\begin{eqnarray}
    \frac{\mathrm{d}\sigma_{s_1s_2}}{\mathrm{d}\Omega} = \frac{\pi\beta\alpha_{em}^2}{s^2}D(s)\left(1+\frac{\alpha}{3}\right)W_{s_1s_2}.
\end{eqnarray}

\normalem

\bibliographystyle{apsrev4-1.bst}
\bibliography{ref.bib}

\end{document}